\documentclass[notitlepage,10pt,aps,prd,twocolumn,amsmath,amssymb,floatfix,superscriptaddress,nofootinbib,preprintnumbers]{revtex4-1}

\usepackage{amsfonts,amssymb,amsmath,graphicx,color,bm,epsfig}
\usepackage[normalem]{ulem}
\usepackage{multirow}
\usepackage{enumerate}
\usepackage[utf8]{inputenc}
\usepackage{hyperref}
\usepackage{braket}
\usepackage{subfigure}
\usepackage{float}
\usepackage[dvipsnames]{xcolor}
\usepackage{soul}
\usepackage{rotating}
\usepackage{physics}
\usepackage{xspace}
\usepackage{bbm}
\usepackage{booktabs}
\usepackage[capitalise]{cleveref}

\hypersetup{
    colorlinks=true,
    linkcolor=red,
    citecolor=blue,
}

\newcommand{\Dth}{\vb*{\Delta \theta}}
\newcommand{\vbth}{\vb*{\theta}}
\newcommand{\ie}{i.e.\xspace}
\newcommand{\vs}{\textit{vs.}\xspace}
\newcommand{\tf}{\textsc{TensorFlow}\xspace}

\newcommand{\cosmosis}{\textsc{CosmoSIS}\xspace}
\newcommand{\multinest}{\textsc{MultiNest}\xspace}

\usepackage{mathtools,xparse}

\RenewDocumentCommand\Pr{sO{}r()}{%
  P
  \begingroup
  \IfBooleanTF{#1}
    {\PrInn*{#3}}
    {\PrInn[#2]{#3}}%
  \endgroup
}

\DeclarePairedDelimiterX\PrInn[1](){%
  \activatebar
  #1%
}

\newcommand{\activatebar}{%
  \begingroup\lccode`~=`|
  \lowercase{\endgroup\def~}{\,\delimsize\vert\,}%
  \mathcode`|=\string"8000
}

\newcommand{\Post}{\mathcal{P}}
\newcommand{\Prior}{\Pi}
\newcommand{\Like}{\mathcal{L}}

\newcommand{\model}{\mathcal{M}}

\allowdisplaybreaks

\AtBeginDocument{
\heavyrulewidth=.08em
\lightrulewidth=.05em
\cmidrulewidth=.03em
\belowrulesep=.65ex
\belowbottomsep=0pt
\aboverulesep=.4ex
\abovetopsep=0pt
\cmidrulesep=\doublerulesep
\cmidrulekern=.5em
\defaultaddspace=.5em
}

\begin{document}

\title{Non-Gaussian estimates of tensions in cosmological parameters}
\author{Marco Raveri}
\affiliation{Center for Particle Cosmology, Department of Physics and Astronomy, University of Pennsylvania, Philadelphia, PA 19104, USA}
\author{Cyrille Doux}
\affiliation{Center for Particle Cosmology, Department of Physics and Astronomy, University of Pennsylvania, Philadelphia, PA 19104, USA}

\begin{abstract}
We discuss how to efficiently and reliably estimate the level of agreement and disagreement on parameter determinations from different experiments, fully taking into account non-Gaussianities in the parameter posteriors.
We develop two families of scalable algorithms that allow us to perform this type of calculations in increasing number of dimensions and for different levels of tensions.
One family of algorithms rely on kernel density estimates of posterior distributions while the other relies on machine learning modeling of the posterior distribution with normalizing flows.
We showcase their effectiveness and accuracy with a set of benchmark examples and find both methods agree with each other and the true tension within $0.5\sigma$ {in difficult cases and generally to $0.2\sigma$} or better.
This allows us to study the level of internal agreement between different measurements of the clustering of cosmological structures from the Dark Energy Survey and their agreement with measurements of the Cosmic Microwave Background from the Planck satellite.
\end{abstract}
\maketitle

\section{Introduction}

Despite the success of the $\Lambda$ Cold Dark Matter ($\Lambda$CDM) model at fitting a vast range of cosmological observations, the emergence of increasingly statistically significant tensions on cosmological parameters has sparked interest in the problem of their precise calculation. 
{In particular, the value of the Hubble constant, $H_0$, inferred from observations of the cosmic microwave background~\cite{Aghanim:2018eyx, Aghanim:2019ame} is lower than that inferred from the distance ladder at low redshifts using type Ia supernovae \cite{Riess:2020fzl}, at 4~to~6$\sigma$, which has prompted searches for hints of new physics \citep[see][for a recent review]{DiValentino:2021izs}. Of less significance, but still puzzling, is the 1~to~2$\sigma$ lower amplitude of structure found by weak lensing surveys~\cite{Abbott:2021bzy,Asgari:2020wuj} compared to that inferred from the CMB.}

Many {tension metrics and numerical} tools have been developed and characterized in the literature (e.g.~\cite{Hobson:2002zf, Marshall:2004zd, Amendola:2012wc, Martin:2014lra, Raveri:2015maa, Seehars:2015qza, Joudaki:2016mvz, Charnock:2017vcd, Lin:2017ikq, Bernal:2018cxc, Raveri:2018wln, Adhikari:2018wnk, Raveri:2019gdp, Handley:2019wlz, Lemos:2019txn, Lin:2019zdn, Park:2019tyw, Miranda:2020lpk, Doux:2020kdz, Lemos:2020jry}).
Most commonly Gaussian approximations at either the data or parameter posterior level need to be made when computing statistical significance.
One open problem is how to precisely compute the statistical significance of a tension in the presence of significant non-Gaussianities in the posterior distributions of cosmological parameters.
While the problem is conceptually straightforward, the large number of dimensions usually involved when comparing data sets of interest in cosmology makes its calculation challenging in practice.

In this paper, we first derive and characterize the posterior distribution of \textit{differences} between parameters determinatined from two, possibly correlated, experiments.
We then discuss two methods to compute the statistical significance of a parameter difference, showcasing their reliability on a set of toy problems and two examples from cosmology.
One class of methods relies on kernel density estimates with fixed and variable data-driven smoothing scales.
Another method relies on machine learning modeling of the posterior distribution with normalizing flows.
Both methods show remarkable accuracy on a set of benchmark examples that cover a wide range of dimensions (from 2 to 30), different known input tension levels (from 1 to 4$\sigma$) and different ways in which non-Gaussianities in the parameter posteriors might arise.
We comment on the advantages and weaknesses of both strategies and develop efficient algorithms for their computation.
Without these algorithms the calculation of statistical significance would be practically impossible in realistic scenarios.
All the methods we discuss have to give the same result and are expected to converge for increasing posterior sample size so their spread can be used as a reliable global error estimate.
Finally, we implement these non-Gaussian tension estimators in \texttt{tensiometer} 
\footnote{\url{https://github.com/mraveri/tensiometer}}, that we use to perform all calculations in this paper and that we make publicly available.

As a first worked example we study the mutual consistency of measurements of galaxy lensing and clustering from the Dark Energy Survey Year~1 data~\cite[DES~Y1,][]{Abbott:2017wau}, finding them in excellent agreement.
We then compute the level of agreement, within $\Lambda$CDM, of DES~Y1 measurements and cosmic microwave background power spectrum measurements from the Planck satellite~\citep{Aghanim:2018eyx, Aghanim:2019ame}, finding them in disagreement at about the $p=0.3\%$ (or $3\sigma$) level.
The former example shows the reliability of these estimates in a case with significant correlations and high dimensionality but low statistical significance, while the latter example showcases reliability in the case of uncorrelated measurements but higher statistical significance.

This paper is structured as follows:
in \cref{Sec:ParamDifferences}, we discuss how to obtain the posterior distribution of parameters differences, calculate their statistical significance and characterize its properties; in \cref{Sec:KdeEstimate}, we show how to efficiently calculate statistical significance with KDE-based methods; in \cref{Sec:FlowEstimate}, we discuss how to obtain statistical significance of parameter differences with normalizing flows; in \cref{Sec:ToyExample}, we discuss a set of benchmark examples that are tailored to gauge the accuracy of both calculation methods; in \cref{Sec:RealExample}, we apply both the KDE and normalizing flows estimators to data from Planck and DES; we discuss the results and conclude in \cref{Sec:Conclusions}.

\section{Parameter differences} \label{Sec:ParamDifferences}

In this section we introduce the distribution of parameter differences and comment on its properties.

We start by defining general notations and writing the posterior probability of parameters $\vbth$, within model $\model$ given data $d$ as
\begin{align} \label{Eq:Posterior}
\Post(\vbth) \equiv \Pr(\vbth|d,\model) = \frac{\Pr(\vbth|\model) \Pr(d|\vbth,\model)}{\Pr(d|\model)} \,,
\end{align}
where we indicate the prior probability as ${\Prior(\vbth) \equiv \Pr(\vbth|\model)}$, the likelihood as ${\Like(\vbth) \equiv \Pr(d| \vbth, \model)}$ and the evidence as $\mathcal{E} \equiv \Pr(d|\model)$. 
We will drop $\model$ hereafter for conciseness and indicate the support of the prior as $V_\Pi$.
{This is defined as the set of parameters where the prior is strictly non-vanishing.}

We now consider two different data sets $d_1$ and $d_2$, labeled here and throughout with $1$ and $2$ subscripts, described by a joint likelihood ${\Like(\vbth)\equiv \Pr(d_1,d_2|\vbth)}$ with shared parameters $\vbth$ and prior $\Prior(\vbth)$. 
We denote ${\Like_1(\vbth)\equiv \Pr(d_1|\vbth)}$ and ${\Post_1(\vbth)\equiv \Pr(\vbth|d_1)}$ the 
joint likelihood/posterior that has been marginalized over $d_2$ (likewise for $d_1$).
The first step to build the parameter difference distribution consists in duplicating all shared parameters, \ie create two copies of the parameter set, $\vbth_1$ and $\vbth_2$, and then choose a prescription for the joint likelihood with duplicated parameters, ${\Like(\vbth_1,\vbth_2)\equiv\Pr(d_1,d_2|\vbth_1,\vbth_2)}$. 
As an example, if the joint likelihood is a multivariate Gaussian distribution over $(d_1, d_2)$ with parameter-independent covariance, a typical choice is to use $\vbth_1$ (respectively $\vbth_2$) in the computation of the mean of $d_1$ ($d_2$). 
In general, however, there is not necessarily a unique solution and different data splits correspond to different aspects that are tested.
We may simply impose the following constraints: (a) ${\Like(\vbth_1=\vbth,\vbth_2=\vbth)=\Like(\vbth)}$ such that the likelihood with duplicated parameters coincides with the joint likelihood on the subset where parameters are equal, and (b) $\Pr(d_1|\vbth_1,\vbth_2)=\Pr(d_1|\vbth_1)$ such that the likelihood for $d_1$ is independent of $\vbth_2$ once marginalized over $d_2$ (and similarly even switching indices 1 and 2). These conditions ensure that, in the case where the data are conditionally independent given $\vbth$, \ie $\Pr(d_1,d_2|\vbth)=\Pr(d_1|\vbth)\Pr(d_2|\vbth)$, then the only choice is to use the product of likelihoods, such that ${\Like(\vbth_1,\vbth_2)=\Like_1(\vbth_1)\Like_2(\vbth_2)}$. We further assume that the joint prior distribution factorizes, $\Pi(\vbth_1, \vbth_2) = \Pi(\vbth_1)\Pi(\vbth_2)$. We can then write the posterior of duplicated parameters, in the general case, as ${\Post(\vbth_1,\vbth_2)\equiv\Pr(\vbth_1,\vbth_2|d_1,d_2)}\propto\Like(\vbth_1,\vbth_2)\Pi(\vbth_1)\Pi(\vbth_2)$.

To obtain the probability density of parameter differences we can change variables and consider $\vb*{\Delta\theta} \equiv \vbth_1 -\vbth_2$. This gives a joint posterior $\Post(\vbth_1, \Dth) = \Post(\vbth_1,\vbth_1-\Dth)$
and we can integrate out the base parameters,
\begin{align} \label{Eq:ParameterDifferencePDF}
\Post(\Dth)  = \int_{V_\pi} \Post(\vbth,\vbth-\Dth) \dd{\vbth} \,.
\end{align}
We distinguish two cases based on whether the two datasets at hand are correlated or not:
\begin{itemize}
\item If the data sets are conditionally independent given parameters $\vbth$, the posteriors $\Post_1$ and $\Post_2$ may be sampled independently, and differences of samples $\vb*{\Delta\theta} \equiv \vbth_1 -\vbth_2$, with ${\vbth_1\sim\Post_1}$ and ${\vbth_2\sim\Post_2}$, are samples of the difference distribution $\Post(\Dth)$, which is given by:
\begin{align} \label{Eq:IndependentParameterDifferencePDF}
\Post(\Dth)  = \int_{V_\pi} \Post_1(\vbth)\Post_2(\vbth-\Dth) \dd{\vbth} \,.
\end{align}
This type of integral is known in the signal processing literature as the cross-correlation of $\Post_1$ and $\Post_2$.
\item If the data sets are correlated, we can obtain samples of $\Post(\Dth)$ by sampling the joint posterior with duplicated parameters and differentiating sample by sample. 
\end{itemize}

Once the density of parameter differences is obtained, one can quantify the probability that there is in fact a parameter shift, given by:
\begin{align} \label{Eq:ParamShiftProbability}
\Delta = \int_{\Post(\Dth)>\Post(\vb{0})} \Post(\Dth) \dd{\Dth} \, ,
\end{align}
which is the posterior mass above the iso-contour of no shift, $\Dth=\vb{0}$.
In the following we indicate the parameter set contained by the iso-contour of no shift as ${\Omega_0 \equiv \{ \Dth \,:\, \Post(\Dth)>\Post(\vb{0})\}}$.

\begin{figure}[!th]
\centering
\includegraphics{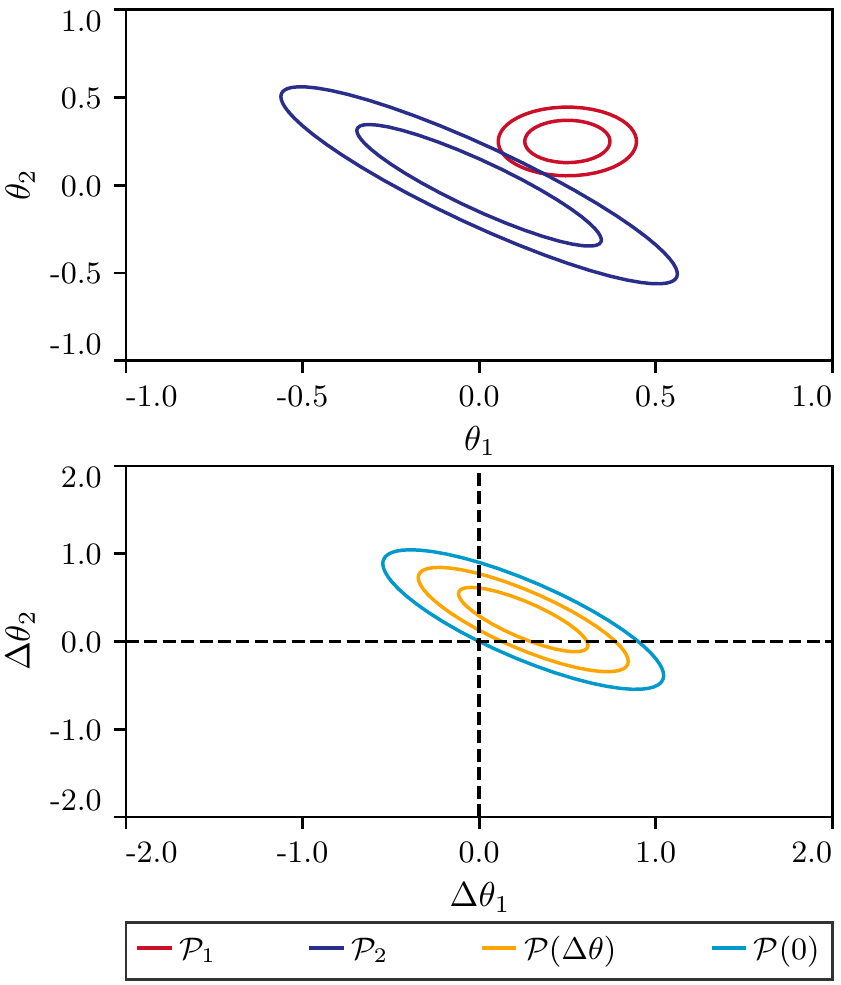}
\caption{ \label{fig:pedagogy}
{
A toy example of how the parameter difference tension calculation proceeds, starting from the two single parameter distributions in the upper panel, to the calculation of the distribution of parameter differences and its iso-contour of zero parameter shift in light blue in the lower panel.
In both panels different lines of the same color show the $68\%$ and $95\%$ C.L. regions of parameter space for different distributions, as shown in legend.
}
}
\end{figure}

Several basic properties can be easily shown from the definition.
\begin{itemize}
\item If the posterior distributions are normalized, then so is the parameter difference distribution. 
\item The support of the parameter difference distribution is the support of the convolution of the characteristic function of $V_\Pi$ with itself.
In particular if the support of the prior is an interval $[a,b]$ then the support of the prior for the parameter difference posterior is $[a-b, b-a]$.
\item The distribution of parameter differences is always continuous at the prior boundary, even if the two single posteriors are not.
Smoothness or continuity of derivatives is not guaranteed.
\item $\Delta$ is symmetric for changes of data sets $1 \leftrightarrow 2$.
\item When datasets are uncorrelated and we have $n_1$ and $n_2$ samples, we can build up to $n_1 n_2$ samples of the parameter difference distribution. If the samples are weighted the weights of the parameter difference samples are the product of the two weights.
\item {If the two posteriors are uncorrelated Gaussian distributions with mean $\theta_1$ and $\theta_2$ and covariance $\mathcal{C}_1$ and $\mathcal{C}_2$ respectively, then the distribution of parameter differences is Gaussian with mean $\theta_1-\theta_2$ and covariance $\mathcal{C}_1 + \mathcal{C}_2$.
If the two parameters are correlated and $\mathcal{C}_{12}$ indicates their covariance, then the covariance of parameter differences is given by $\mathcal{C}_1 + \mathcal{C}_2 -\mathcal{C}_{12} -\mathcal{C}_{12}^T$ and the mean is unchanged.}
\end{itemize}

{In \cref{fig:pedagogy} we show an example of how the calculation of the parameter difference tension estimator proceeds.
We start in the upper panel showing two Gaussian posterior distributions, with different parameter degeneracies and different means. We then obtain and show in the lower panel, the distribution of parameter differences, as the convolution of the two single distributions, as in \cref{Eq:IndependentParameterDifferencePDF}.
Since the two initial distributions are Gaussian the distribution of parameter differences is Gaussian as well.
The support of the prior for the two separate posteriors is the $[-1,1]$ square, hence the support of the prior of parameter differences is the $[-2,2]$ square. 
We then find the iso-contour that touches the origin. 
Integrating the posterior inside that iso-contour indicates that it contains $0.995$ of the posterior mass, which is the estimate of the probability that the parameters are actually different. 
}

In the following sections we comment on other properties of parameter differences that are worth further discussion.

\subsection{Unconstrained parameters} \label{Sec:FlatPosterior}

A couple of general results can be shown from the definition of the parameter difference distribution and help its characterization in cases where data do not constrain some of the parameters of the model.

The first consists in showing that parameters that are prior constrained for both data sets can never contribute a shift.
If there is no data information on a parameter then the likelihood of both experiments is constant and the parameter shift integral becomes:
\begin{align}
P(\Dth) \propto \int \Pi(\vbth)\Pi(\vbth-\Dth) \dd{\vbth}
\end{align}
up to an irrelevant normalization constant.
In signal processing literature this integral is called the autocorrelation of $\Pi$.
By Cauchy inequality this is maximum for $\Dth=\vb{0}$ so that the posterior can have no mass above it.

The second noticeable property is that parameters constrained by flat priors for one data set but not the other can never contribute a shift.
If the prior is flat across the support and that the second data set provides no constraint (\ie has a constant likelihood), we have:
\begin{align} \label{Eq:ShiftFlatPrior}
P(\Dth) \propto \int \Prior(\vbth)\Like(\vbth) \Prior(\vbth-\Dth) \dd{\vbth},
\end{align}
up to an irrelevant normalization constant.
\Cref{Eq:ShiftFlatPrior} is integrating the likelihood over the intersection between the prior volume and the shifted prior volume.
A non vanishing $\Dth$ reduces the integration volume of a positive function and hence reduces the integral.
This means that \cref{Eq:ShiftFlatPrior} is maximum at zero shift.

As a consequence of this second property non-shared parameters between two experiments can never contribute any shift and can be pre-marginalized over before doing any parameter shift calculation.

\subsection{Properties under changes of parameters} \label{Sec:Reparametrizations}

Since \cref{Eq:ParamShiftProbability} is the integral of a probability density it is invariant under reparametrizations.
However, when working directly with transformations of parameter differences some care must be exercised.

We consider a change of variables $\vb{z}=\vb{T}^{-1}(\Dth)$ with an invertible, differentiable transformation $\vb{T}$. The probability density function of $\vb{z}$ is given by:
\begin{equation} \label{eq:jac}
\Pr(\vb{z})=\Post(\Dth) \abs{\det \grad \vb{T}} \,,
\end{equation}
computed with the determinant of the Jacobian of $\vb{T}$. The shift probability is then given by:
\begin{equation}
\Delta = \int_{\Omega_0} \Post(\Dth) \dd{\Dth}
    = \int_{\vb{T}^{-1}({\Omega_0})} \Pr(\vb{z}) \dd{\vb{z}}.
\end{equation}
where both the posterior and integration measures change by the same factor that then cancels; but the domain of integration is modified to the pre-image of $\Omega_0$, which may not necessarily correspond to the zero iso-contour of 
$\Pr(\vb{z})$.

\begin{figure}
\centering
\includegraphics{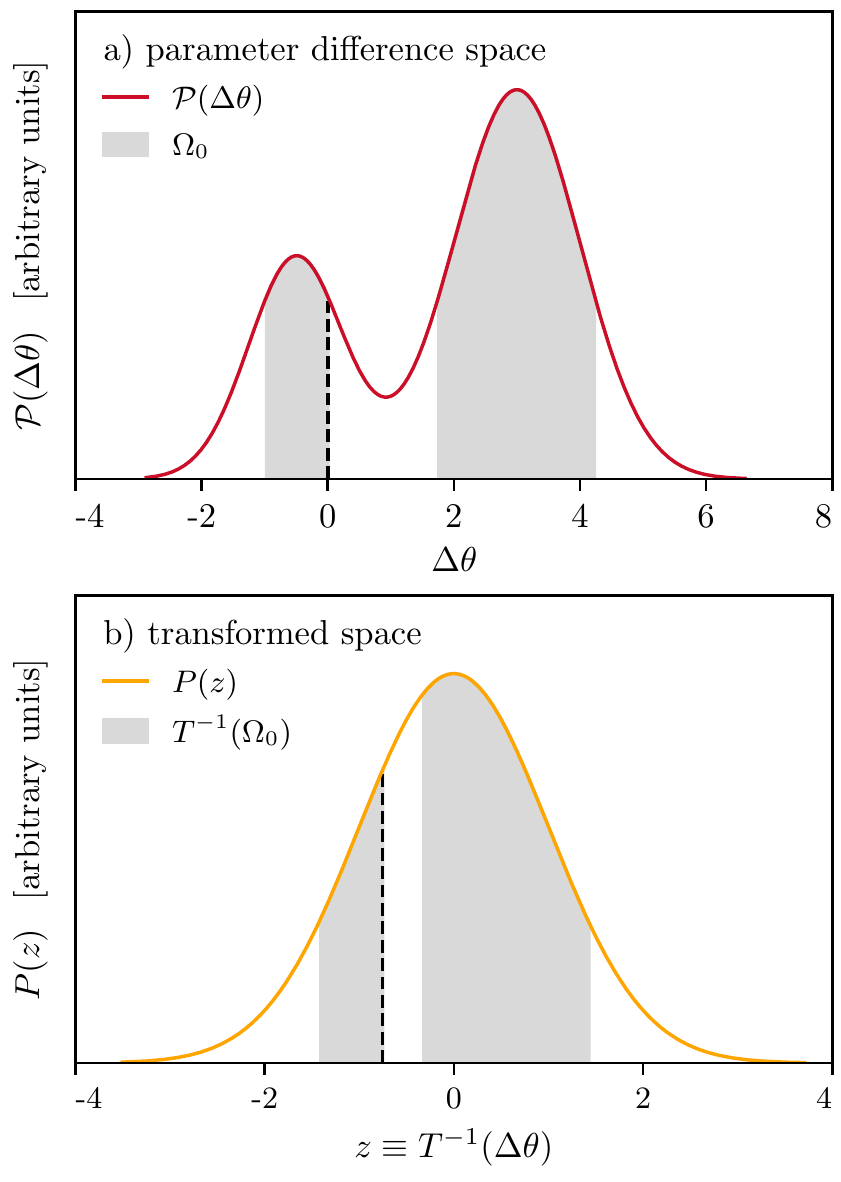}
\caption{ \label{fig:1d_gmm}
A toy example of a gaussianizing transformation showing the mapping of the iso-contour at a given point.
}
\end{figure}

In~\cref{fig:1d_gmm}, we consider the one-dimensional case where $\Post(\Dth)$ is a mixture of gaussian distributions with two components. 
We apply the gaussianizing transformation given by the successive application of the cumulative distribution function (CDF) of the gaussian mixture and the inverse CDF of the standard normal distribution. 
The resulting transformation is highly non-linear.
The top panel shows the posterior $\Post(\Dth)$ and the bottom panel shows $\Pr(\vb{z})$. 
The iso-contour at zero-shift, $\Omega_0$, shown in gray in the top panel, has two connected components. 
Its pre-image, shown in gray in the bottom panel, also has two components, and therefore $\vb{T}^{-1}({\Omega_0})$ clearly does not correspond to an iso-contour of the Gaussian distribution $\Pr(\vb{z})$.
Note, however, that the posterior mass in the two gray areas is conserved and the same in the two cases.

\section{KDE estimate of parameter differences} \label{Sec:KdeEstimate}

In this section we discuss how to compute the integral in \cref{Eq:ParamShiftProbability} with Kernel Density Estimates (KDE) of the parameter difference posterior.
For an overview of KDE techniques we refer the reader to~\cite{wand1994kernel} and~\cite{chacon2018multivariate}.

As we have highlighted in \cref{Sec:ParamDifferences} when building the parameter difference chain we do not have likelihood values for samples.
In this section we comment on how to obtain likelihood values through KDE.
In this section only we use the shorthand notation of $P(\vb{x}) \equiv \Post(\Dth)$.

Given a kernel smoothing function, $K(\vb{x})$, that is positive, smooth, spherically symmetric and normalized we consider a bandwidth matrix $\vb{H}$ which we assume to be symmetric and positive definite.
The bandwidth matrix scales the kernel coordinates $K_{\vb{H}}(\vb{x}) \equiv |\vb{H}|^{1/2} K(\vb{H}^{-1/2} \vb{x})$ and controls the extent of smoothing in different directions.
In practice we always assume that the smoothing kernel $K_{\vb{H}}(\vb{x})$ is Gaussian.

Given our $n$ weighted samples, $\vb{X}_i$, of $P$, we can compute its KDE estimate as:
\begin{align} \label{Eq:KDEDifferenceProbability}
\hat{P}_{\vb{H}}(\vb{x}) \equiv \frac{1}{w_{\rm tot}}\sum_{i=1}^{n} w_i \, K_{\vb{H}}(\vb{x} -\vb{X}_i) \;,
\end{align}
where $w_i$ denotes the weights of the samples and $w_{\rm tot} \equiv \sum_{i=1}^{n} w_i$.
In abstract terms our problem consists in estimating the posterior mass inside the iso-contour that touches a particular point, $\vb{x}_0$, whose KDE estimate we indicate with:
\begin{align} \label{Eq:KDEParamShiftProbability}
\hat{\Delta} = \int_{\hat{P}(\vb{x})>\hat{P}(\vb{x}_0)} \hat{P}(\vb{x}) \dd{\vb{x}} \,.
\end{align}
To the best of our knowledge this problem has not been specifically addressed in KDE literature.

We notice that since the distribution of parameter differences is always continuous at the boundary we do not need to be concerned about correcting for KDE biases at the boundary.

When we consider high dimensional problems it is useful to slightly modify the KDE estimator to have adaptive bandwidth. This leads to the sample point density estimators:
\begin{align}
\hat{P}_{\rm SP}(\vb{x}) \equiv \frac{1}{w_{\rm tot}}\sum_{i=1}^{n} w_i \, K_{\vb{H}_i}(\vb{x} -\vb{X}_i) \;,
\end{align}
which considers a different bandwidth matrix for every sampled point $\vb{H}_i$.
As discussed in~\cite{terrell1992} we can consider different strategies to fix the sampled point bandwidth.
In particular we test the case where $\vb{H}_i = \delta_{(k)}(\vb{X}_i)^2 \vb{I}$, where $\delta_{(k)}$ is the distance to the $k$ nearest neighbor of a given point.
Note that for sample point density estimators there is no smoothing parameter to optimize.
We tested other variable smoothing scale choices, in particular nearest neighbor ellipsoidal smoothing, but find that they never exceed the performances of the simpler nearest neighbor distance smoothing scale that we use.

\subsection{Monte-Carlo algorithms} \label{Sec:KDEMonteCarlo}

Since we have samples from $P$ we can perform the integration of \cref{Eq:KDEParamShiftProbability} as a Monte Carlo (MC) volume integral:
\begin{align} \label{Eq:KDEMCMCParamShiftProbability}
\hat{\Delta} = \frac{1}{w_{\rm tot}} \sum_{i=1}^{n} w_i S( \hat{P}(\vb{X}_i)-\hat{P}(\vb{x}_0) )
\end{align}
where $S(x)$ is the Heaviside step function that is unity when $x>0$ and zero otherwise.
We note that we have here assumed that the samples from the parameter difference chain are uncorrelated.
This is generally easily achieved by suitably thinning the input parameter samples in order to obtain uncorrelated samples.
Since we are counting samples above a threshold, the confidence intervals for the result can be estimated from the binomial distribution, through the Clopper–Pearson formula for the sum of weights above threshold and total sum of weights.
We notice that the Clopper–Pearson formula applies to non-integer values of trials and successes so we can use it with non-integer weights.

We can compute $\hat{\Delta}$ with a {\it brute force} algorithm that iterates through the samples, computes KDE probability of a sample, compares its value to the pre-computed reference and counts the sample or not.
The drawback of this algorithm is that it clearly requires $\order{n^2}$ operations making it unfeasible for high statistically significant tensions that require a large number of samples to be accurately estimated. 

For high statistically significant tension, we develop an ad-hoc algorithm that we call {\it neighbor elimination}.
We notice that \cref{Eq:KDEMCMCParamShiftProbability} involves a threshold crossing problem, whether $\hat{P}(\vb{X}_i)>\hat{P}(\vb{x}_0)$ or not, but does not rely on the actual value of $\hat{P}(\vb{X}_i)$.
Each of the $\hat{P}(\vb{X}_i)$ is computed from \cref{Eq:KDEDifferenceProbability} and is a sum of positive quantities that decrease with distance from $\vb{X}_i$, since the kernel is positive.
Then, after pre-computing $\hat{P}(\vb{x}_0)$, we start by organizing samples in a k-d tree, which costs $\order{n \log n}$ operations.
We then iterate through the samples and query the tree for the nearest neighbor of each point, which globally costs $\order{n \log n}$ operations.
We build the sum in \cref{Eq:KDEDifferenceProbability} going through the nearest samples and since the kernel is positive we can discard a point in the tree as soon as it crosses the threshold of $\hat{P}(\vb{x}_0)$.
For a large tension, where $\hat{P}(\vb{x}_0)$ is small and in the tails of the distribution, it only takes a few nearest neighbors to be able to discard a point. If the vast majority of points are discarded after a search of $\alpha$ nearest neighbor then the calculation of \cref{Eq:KDEParamShiftProbability} only requires $\order{(\alpha + 1) n \log n}$ operations. Notice that in the worst case scenario, where all points are above threshold, this algorithm requires $\order{n^2 \log n}$ operations so it is only suited for the calculation of statistical significance of large tensions.

Problems in one or two dimensions are, in this respect, special, since KDE evaluation can be performed with Fast Fourier Transforms (FFT) on a discrete grid with $n_{\rm grid}$ points that is then interpolated. This decreases the cost of operations to be globally $\order{n_{\rm grid} n \log n}$ and makes it clearly faster than {\it neighbor elimination} below three dimensions. For and above dimension three, we expect the {\it neighbor elimination} algorithm to outperform FFT methods mostly because of the increasing size of the grids involved in the FFT calculation.

Notice that the {\it brute force} and the {\it neighbor elimination} algorithms have to agree exactly on the result while the FFT algorithm can differ with dependence on the size of the grid.

\subsection{KDE bandwidth selection} \label{Sec:KDEBandSelection}

The last ingredient we need to estimate \cref{Eq:ParamShiftProbability} with KDE is a selection of the bandwidth matrix $\vb{H}$.
We refer the reader to~\cite{chacon2018multivariate} for a comprehensive discussion of multidimensional band selectors. 
Here we review the main ideas and outline the strategies that we compare in the next sections.

Although we phrase the following in general terms, in practice, we always work with pre-whitened samples, scaling all samples as $\tilde{\vb{x}} = \vb{\Sigma}^{-1/2}\vb{x}$, where $\vb{\Sigma}$ is the sample parameter covariance. Doing this ensures that sample variance is the identity matrix. This cannot generally be done in all applications and we exploit here explicitly the independence of our resulting integral from the choice of coordinates.

The first family of band selectors that we consider aim at minimizing the mean integrated square error (MISE):
\begin{align} \label{Eq:KDEMISE}
{\rm MISE}(f) \equiv \int \mathbb{E} \qty[(f(\vb{x})-\hat{f}(\vb{x}))^2] \dd{\vb{x}} 
\end{align}
where $\mathbb{E}$ denotes expectation over independent and identically distributed samples realizations.
Minimizing this quantity is a very common approach as it seeks to minimize differences between the true distribution and its KDE approximation over the whole domain, resulting in a KDE estimate that is overall good for general purposes.

As noted in the two seminal papers~\cite{Rosenblatt:1956:RSN, parzen1962}, and can be verified by direct calculation, \cref{Eq:KDEMISE} can be written exactly in terms of convolutions between the true distribution $f$ and the smoothing kernel $K_{\vb{H}}$:
\begin{align} \label{Eq:exactKDEMISE}
{\rm MISE}(f) =& R(K_{\vb{H}}*f) -2\int (K_{\vb{H}}*f)(\vb{x}) f(\vb{x}) \dd{\vb{x}} +R(f) \nonumber \\
& +n_{\rm eff}^{-1}\,|\vb{H}|^{-1/2} \, R(K) -n_{\rm eff}^{-1}\, R(K_{\vb{H}}*f)
\end{align}
where $n_{\rm eff} \equiv (\sum_i w_i)^2 / (\sum_i w_i^2)$, $R(f) \equiv \int f^2(\vb{x}) \dd{\vb{x}}$ and we indicate convolutions with $f*g \equiv \int f(\vb{x}-\vb{y}) g(\vb{y}) \dd{\vb{y}}$.

To make progress we assume that both the kernel and the target distributions are Gaussian,
with covariance $\vb{H}$ and $\vb{\Sigma}$ respectively, that allows to write \cref{Eq:exactKDEMISE} as
\begin{align} \label{Eq:MinMISE}
{\rm MISE}(\vb{H}) =& (2\pi)^{-d/2} \bigg[ |2\vb{H} + 2\vb{\Sigma}|^{-1/2} -2|\vb{H} + 2\vb{\Sigma}|^{-1/2} \nonumber \\
& \hspace{1.5cm} +|2\vb{\Sigma}|^{-1/2} \bigg] \nonumber \\
& + n_{\rm eff}^{-1}(2\pi)^{-d/2} \bigg[ |2\vb{H}|^{-1/2} -|2\vb{H} + 2\vb{\Sigma}|^{-1/2} \bigg] \,. \nonumber \\
\end{align}
The matrix $\vb{H}_{\rm MISE}$ that minimizes \cref{Eq:MinMISE} cannot be expressed analytically.
Hence our strategy is to minimize \cref{Eq:MinMISE} numerically over the space of positive symmetric matrices~\cite{2019arXiv190600587B}. 

On the other hand if we consider the asymptotic expansion of \cref{Eq:exactKDEMISE} for large $n$ and small $\vb{H}$, usually referred to as 
asymptotic MISE (AMISE), we can solve the corresponding minimization problem exactly in the Gaussian case by:
\begin{align}
\vb{H}_{\rm AMISE} = \left( \frac{4}{(d+2)n_{\rm eff}} \right)^{\frac{2}{d+4}} \vb{\Sigma} \,.
\end{align}
This is known as the rule of thumb selector that we test and generally use as the starting guess for numerical minimization of MISE.

We test other possible choices for bandwidth selectors, including the maximum bandwidth and cross validation.
In practice we find that their performances do not exceed the MISE/AMISE ones at the specific problem we are considering.

\section{Normalizing flow estimate of parameter differences} \label{Sec:FlowEstimate}

We now turn to a radically different method to compute the probability of no shift, $\Delta$, based on building an analytic approximation of $\Post(\Dth)$. This method relies on normalizing flows (NF), that were introduced as generative models with tractable likelihoods, \ie analytic approximation to the likelihood enabling fast sampling and inference \cite{2017arXiv170507057P,2018arXiv180703039K,2018arXiv181001367G,Kobyzev:2019fa,2020arXiv200301941M}. NF have recently found applications in astrophysics, both for modelling \cite{2020arXiv200600615R} and Bayesian inference tasks \cite{2020PhRvD.102j3507D,2020MNRAS.496..328M,2020MNRAS.tmp.3445J}.

\subsection{Approximating difference distributions with normalizing flows} \label{Sec:NFmethod}

In this approach, one introduces a bijective, differentiable mapping $\vb{T}_{\vb*{\varphi}}$ parametrized by neural networks with weights $\vb*{\varphi}$, between $D$-dimensional input data $\vb{x}$ with distribution $P(\vb{x})$ and abstract variables $\vb{z}=\vb{T}_{\vb*{\varphi}}\qty(\vb{x})$. The mapping is optimized over parameters $\vb*{\varphi}$ with (stochastic) gradient descent to maximize the log-probability of $\vb{z}$ under a simple distribution, typically the standard $D$-dimensional normal distribution $\mathcal{N}(\vb{0},\vb{I})$, with density~$\varphi_D$. One finally approximates the data distribution $P$ with ${P(\vb{x})\approx q(\vb{x}) = \varphi_D(\vb{z}) \abs{\det \grad_{\vb{x}}{\vb{T}_{\vb*{\varphi}}}}}$, using the determinant of the Jacobian of the transformation, as explained in \cref{Sec:Reparametrizations}. The density of the approximated distribution $q(\vb{x})$ can be computed analytically using automatic differentiation. One can also easily sample from $q$, by sampling from the $D$-dimensional standard Gaussian and transforming samples with the inverse mapping, $\vb{x}=\vb{T}_{\vb*{\varphi}}^{-1}\qty(\vb{z})$.

For the problem at hand in this work, we approximate the parameter difference distribution $\Post(\Dth)$ with $q(\Dth) = \varphi_D(\vb{z}) \abs{\det \grad_{\Dth}{\vb{T}_{\vb*{\varphi}}}}$, where $\vb{z}=\vb{T}_{\vb*{\varphi}}(\Dth)$ and $\vb{T}_{\vb*{\varphi}}$ has been optimized to gaussianize $\vb{z}$, as in \cref{Sec:Reparametrizations}. This is obtained by minimizing the loss function given by ${L(\Dth,\vb*{\varphi})=-\log\varphi_D(\vb{T}_{\vb*{\varphi}}(\Dth))}$. In order to estimate the shift probability $\Delta$, we sample ${\Dth\sim q}$,  compute the probability density of those samples, $q\qty(\Dth)$, and compare the results to $q\qty(\Dth=\vb{0})$, computed the same way. The fraction of samples for which $q(\Dth)>q(\vb{0})$ provides a Monte-Carlo estimate of the shift probability $\Delta$, with an error that can be made arbitrary low by generating more samples and estimated with the Clopper-Pearson formula. The accuracy of the estimate of $\Delta$ is therefore limited by the the accuracy of the normalizing flow model, rather than the number of samples drawn during this final step.

This technique transforms the problem of estimating $\Delta$ into an optimization problem. This means that it is not guaranteed to converge to the true value and instead, the result will depend on the choice of NF and neural network architecture, as well as drawbacks inherent to neural networks such as initialization, underfitting and overfitting. In particular, if the tension is strong and the no-shift point is far from the bulk of samples, the approximation to $\Post(\vb{0})$ may be a poor one. Nonetheless, the computation---which involves training a neural network---now benefits from a large initial sample and suffers less than KDE from dimensionality \citep{bengio:2007}. Moreover, we find it efficient for small to mild tensions, which are interesting to characterize precisely, and thus this methods complements the KDE method presented in the previous section.

\subsection{Normalizing flows implementation} \label{Sec:NFimpl}
In this work, we use Masked Autoregressive Flows \citep[MAF,][]{2017arXiv170507057P} to approximate the parameter difference distribution $\Post(\Dth)$. MAFs are composed of a series of Masked Autoencorders for Density Estimation \cite[MADE,][]{2015arXiv150203509G} that each implement autoregressive transformations of the input variables encoded by a single neural network.
If the input variable is ${\vb{x}=\qty(x_1,\dots,x_n)}$, the output $\vb{y}$ of an autoregressive transformation has components ${y_i=\mu(\vb{x}_{1:i-1})+\sigma(\vb{x}_{1:i-1})x_i}$, where $\mu$ and $\sigma$ are parametrized by unconstrained neural networks that receive masked inputs, ${\vb{x}_{1:i-1}=\qty(x_1,\dots,x_{i-1},0,\dots,0)}$.
MAFs allow for estimation of more complex distributions by stacking such transformations and introducing random permutations of the components between them.

In practice, we first compute the mean $\expval{\Dth}$ and covariance $\bm{\Sigma}$ of the difference samples, $\Dth$, to shift and rescale parameters, $\Dth'=\bm{\Sigma}^{-1/2} (\Dth - \expval{\Dth}) $, thus defining a bijective mapping into a pre-whitened (but still non-Gaussian) space. We then apply MAFs to learn the mapping to the fully gaussianized space and apply chain rules to approximate the difference parameter distribution $\Post$. We find that MAFs are generally sufficiently flexible for typical parameter difference distributions, requiring only few hundreds to few thousands of trainable weights in the examples considered below.

Our code relies on the implementation of MAFs within the \texttt{tensorflow-probability} package \cite[TFP,][]{2017arXiv171110604D}, which benefits from the automatic differentiation of the \tf\footnote{\url{https://www.tensorflow.org/}} framework to compute the Jacobian of the transformation and other derivatives needed for gradient descent. We provide utility functions to quickly instantiate a default (but easily tweakable) architecture of MAFs composed of $2d$ stacked MADEs, each with two dense layers of $2d$ nodes each with $\sinh$ activation functions, and with random permutations of variables between MADEs (where $d$ is the dimension of the parameter difference space). We found this architecture to work well for all the examples detailed below. However, the code is not restricted to using MAFs and can accomodate any user-provided normalizing flow implemented as a TFP \texttt{Bijector} object. We also found that the uniform Glorot initializer (of the weights and biases of MAFs) provided a good balance between learning speed and stability, although the user can also easily opt for any other initializer.
We provide plotting utilities to monitor the evolution of a number of metrics during training (see \cref{Fig:ExTraining}), in addition to the standard training and validation losses, to assert the quality of the approximation to the parameter distribution $\Post$. In particular, since NFs learn a mapping to the standard multivariate Gaussian, we compute the squared norms of the transformed samples and verify that their distribution converges to a $\chi^2_d$ distribution, e.g. with a Kolmogorov-Smirnov test.

\section{Benchmark examples} \label{Sec:ToyExample}

In this section, we perform numerical experiments on benchmark examples designed to test and compare our methods in limit cases. 

We first consider, in \cref{Sec:GaussianExample}, examples where the difference parameter distributions are Gaussian in an increasing number of dimensions. In these examples, the input tension is known analytically.
We then consider, in \cref{Sec:NonGaussianExample,Sec:InformativePriorExample,Sec:MultiModalExample}, two-dimensional models where the two posteriors, $\Post_1$ and $\Post_2$, have analytic probability densities that are strongly non-Gaussian.
The density of the difference distribution, $\Post(\Dth)$, can be computed on a dense grid and is given by the convolution of the posterior densities evaluated over dense grids covering the prior volume $V_\Prior$. 
We then compute the expected probability of no shift by numerical integration of $\Post(\Dth)$ over $\Omega_0$.
In all cases, we compare the probability of no shift obtained with the numerical integration to the one obtained with the KDE and NF methods as well as the Gaussian approximation.

For the Gaussian approximation we compute parameter shifts both in standard form and in update form, as in~\cite{Raveri:2018wln}.
The standard form of parameter shift, $Q_{\rm DM}$, requires to compute:
\begin{align}
Q_{\rm DM} \equiv (\vb*{\theta}_{p1} - \vb*{\theta}_{p2})(\vb*{\mathcal{C}}_{p1} + \vb*{\mathcal{C}}_{p2})^{-1}(\vb*{\theta}_{p1} - \vb*{\theta}_{p2})^T
\end{align}
where $\vb*{\theta}_{p1}$/$\vb*{\theta}_{p2}$ and $\vb*{\mathcal{C}}_{p1}$/$\vb*{\mathcal{C}}_{p2}$ denote the posterior mean parameters and posterior covariance respectively for data set $1$/$2$.
$Q_{\rm DM} $ is $\chi^2$ distributed with ${\rm rank}(\vb*{\mathcal{C}}_{p1} + \vb*{\mathcal{C}}_{p2})$ degrees of freedom.
The update parameter shifts estimator is defined as:
\begin{align}
Q_{\rm UDM} \equiv (\vb*{\theta}_{p1} - \vb*{\theta}_{p12})(\vb*{\mathcal{C}}_{p1} - \vb*{\mathcal{C}}_{p12})^{-1}(\vb*{\theta}_{p1} - \vb*{\theta}_{p12})^T
\end{align}
where $\vb*{\theta}_{p12}$ and $\vb*{\mathcal{C}}_{p12}$ are the joint posterior mean parameters and covariance.
$Q_{\rm UDM}$ is $\chi^2$ distributed with ${\rm rank}(\vb*{\mathcal{C}}_{p1} - \vb*{\mathcal{C}}_{p12})$ degrees of freedom.
We refer the reader to~\cite{Raveri:2018wln} for the details of these estimators and their computation.
Two noteworthy properties of these estimators, in this context, are that they have to coincide for Gaussian distributions with uninformative priors and that the $Q_{\rm UDM}$ estimator is symmetric when swapping data sets $1\leftrightarrow 2$.
In the non-Gaussian case though, the update estimator allows us to use the most Gaussian of the two distributions at hand hence providing mitigation of non-Gaussian features.

For KDE methods we avoid using the FFT algorithm in two dimensions since we want to focus on the properties of the Monte-Carlo algorithm. 
We also use the {\it neighbor elimination} algorithm in all calculations since we find that it always exceeds the performances of the {\it brute-force} algorithm.
We focus on the dependence of the KDE result on the number of samples that are used in the calculation and different bandwidth choices.

For the NF method, we focus on choices of hyperparameters, the impact of the number of samples used to train the model and model uncertainties.
As previously explained, the stochasticity in the initialization of the neural networks and that of the gradient descent during training introduce extra variance in the estimated shift probability which is difficult to evaluate. Therefore, for each example, we perform ten runs and report the uncertainty of the results. 
For Gaussian examples in dimension $d$, we use MAFs made of $\min(d,12)$ stacked MADEs, each with two layers of $2d$ units. We focus on the dependence of the shift probability estimates and their error as a function of the number of samples. 
For two-dimensional examples, we focus on hyperparameters and compare results for MAFs made of either four or eight stacked MADEs, each with two hidden layers of eight units (for a total of 528/1056 parameters), with fixed training samples with a size of $10^5$.
Additionally, the learning rate needs some tuning for these non-Gaussian examples: if it is too large, it may prevent convergence, while if it is too small, the optimizer might fall into a local minimum far from the global one. Therefore, we repeat these tests for learning rates of $10^{-2}$, $10^{-3}$ and $10^{-4}$. For all tests, the estimated shift probability is obtained by sampling the approximate posterior $q$ until the confidence interval on the result, estimated from the Clopper-Pearson formula, has a length smaller than 0.05.

In the following, we always report results as effective number of standard deviations.
Given an event of probability $P$, it is given by
\begin{align} \label{Eq:EffectiveSigmas}
n_\sigma \equiv \sqrt{2}{\rm Erf}^{-1}(P).
\end{align}
This corresponds to the number of standard deviations that an event with the same probability would have had were it drawn from a Gaussian distribution.
This does not imply Gaussianity of the underlying statistics and should be regarded as a logarithmic scale for probabilities.
Note that, when gauging accuracy of estimates, requirements of a precise determination of $n_\sigma$ is a much stricter requirement than a threshold on absolute accuracy of the $P$ value.

\subsection{High-dimensional Gaussian} \label{Sec:GaussianExample}

We start by considering a Gaussian distribution in an increasing number of dimensions. 
Since \cref{Eq:ParameterDifferencePDF} preserves Gaussianity, if the two original distributions are Gaussian, so is the parameter difference distribution.
We then generate a random parameter difference distribution by generating a random mean vector on the sphere, a random covariance, and rescaling the mean vector length to match the desired benchmark tension.

\subsubsection{Results with KDE}

\begin{figure}
\centering
\includegraphics{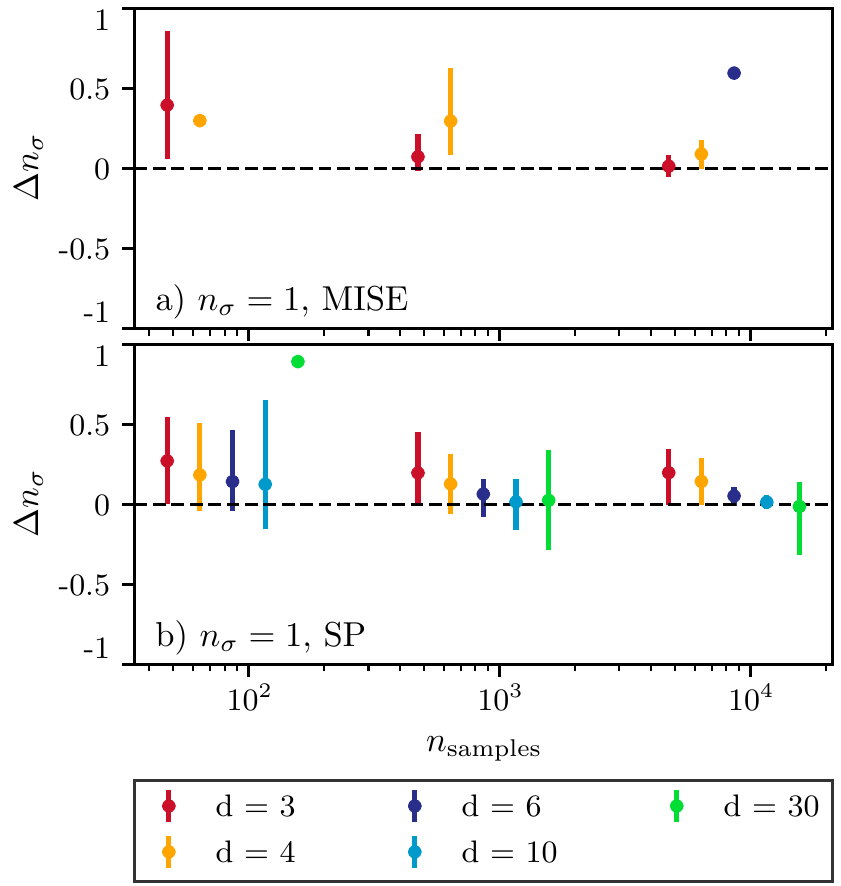}\\
\caption{\label{Fig:GaussianKDE1}
The mean error {$\Delta n_\sigma$} on the determination of a $1 \sigma$ tension {($n_\sigma=1$) from KDE estimates} in increasing number of dimensions as a function of number of samples {$n_{\rm samples}$} and dimension {$d$}.
Different panels show different {KDE} algorithms while different colors show different number of dimensions, as reported in legends. {Error bars are obtained by drawing multiple samples for each case. Note that for high dimensions, samples become sparser and the MISE estimate cannot be computed, which explains missing points.}
}
\end{figure}

\begin{figure}
\centering
\includegraphics{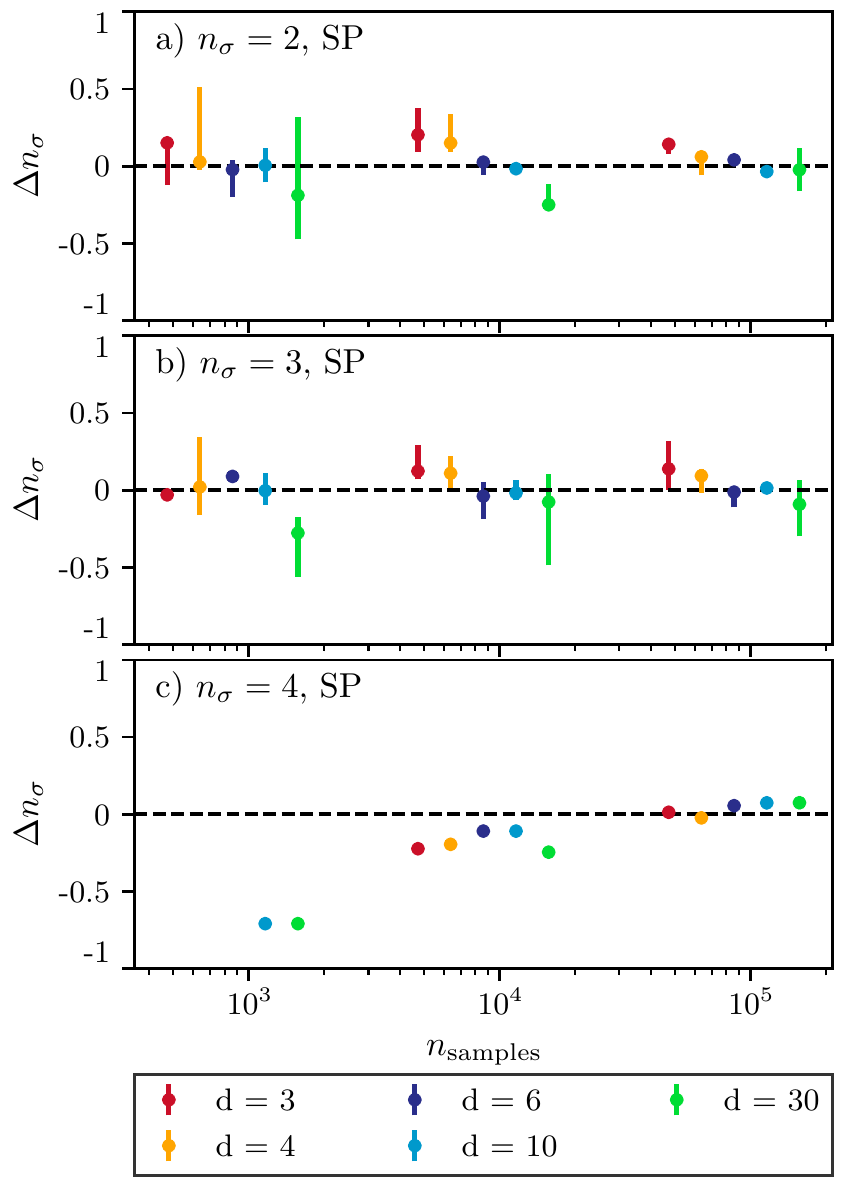}\\
\caption{\label{Fig:GaussianKDE2}
The mean error {$\Delta n_\sigma$} on the determination of $2 \sigma$, $3 \sigma$ and $4 \sigma$ tensions {from the KDE SP estimate} in increasing number of dimensions as a function of number of samples {$n_{\rm samples}$} and dimension {$d$}.
Different panels show different input tension levels while different colors show different number of dimensions, as reported in legends. {For high tensions, small samples lead to highly biased results that are not visible in the lower plot.}
}
\end{figure}

In \cref{Fig:GaussianKDE1} {we show the error ${\Delta n_\sigma}$ on the determination of a $1\sigma$ tension ($n_\sigma=1$) using KDE estimates, and} we compare the KDE estimate with the MISE smoothing scale {to} the KDE SP estimate.
For each example, we draw several batches of {increasingly larger} samples {from the random parameter difference distribution in order} to evaluate the bias of the estimator and its variance{, as a function of number of samples $n_{\rm samples}$ and dimension $d$}.
We find that the MISE estimate converges more rapidly than the SP estimate with the number of samples.
As it is known, however, its bias is, in all cases, larger than its variance.
However, when increasing dimension at fixed sample size, samples become increasingly sparse and the MISE estimate cannot be computed, while the SP estimator retains workable accuracy in all dimensions, even with a relatively small number of samples.

This trend is confirmed in \cref{Fig:GaussianKDE2} where we show examples with higher tensions.
Increasingly, samples around the zero shift contour become sparse so the MISE estimate would require a very large number of samples to be meaningfully computed.
On the other hand the SP estimate shows remarkable performances with error {${\Delta n_\sigma}$} rarely exceeding $0.5\sigma$ independently of dimension and sample size.
We note that the performances of the SP estimator improve as dimension increases and especially in high dimensions the variance of the estimator over pools of samples is of the same order as its bias.

{For large number of samples the variance of the estimator is so small that it cannot be seen in the figure. Some results for low number of samples also show no variance or largely underestimated variance because they are generally borderline feasible and the calculation fails many times because there is no sample outside the iso-contour of no shift. }

In all cases we note that the error scaling is worse than what we would expect from binomial trials.
The reason why this happens is that the zero shift iso-contour which is the decision boundary in the binomial trial for each sample is inferred from the samples. Sampling uncertainty on its estimate then contribute to the total error budget.

We test several different algorithms and reach to qualitatively similar conclusions.
The AMISE bandwidth leads to similar results as the MISE one, while cross-validation band estimators generally lead to a smaller smoothing scale, which then requires more samples to converge.
Adaptive KDE with ellipsoidal smoothing scale show performances similar to the SP estimator.

\subsubsection{Results with normalizing flows}

\begin{figure}
\centering
\includegraphics{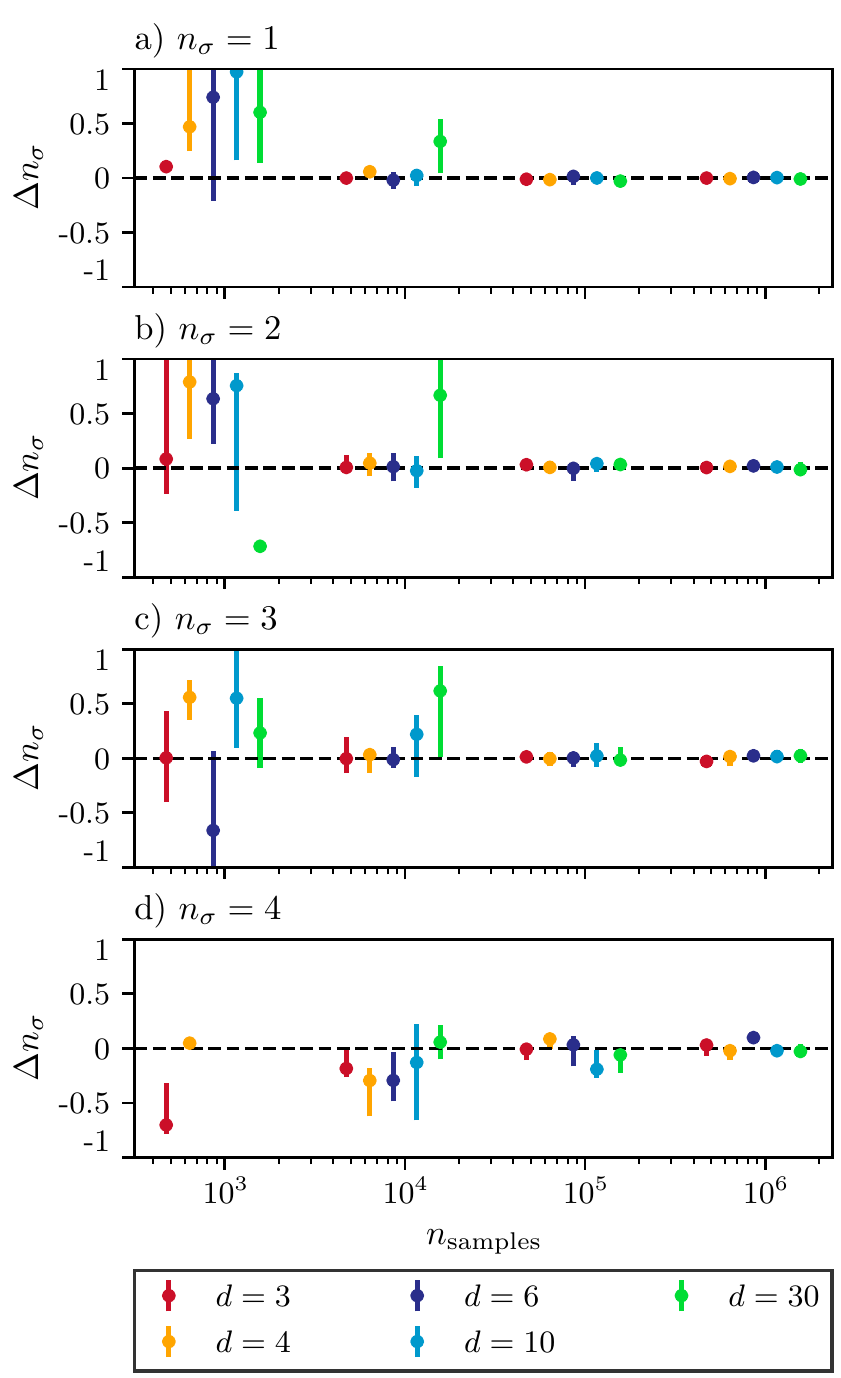}\\
\caption{\label{Fig:GaussianNF}
{Similar to \cref{Fig:GaussianKDE1,Fig:GaussianKDE2} for the NF estimate.}
}
\end{figure}

In \cref{Fig:GaussianNF}, we show
{the bias and variance of the NF estimate}
over ten runs, for increasing levels of tension and increasing dimensions, as a function of the number of training samples. In order to obtain a fair comparison, we fix the number of epochs during training to 100, each with 100 batches per epoch (possibly using samples several times per epoch). This sets the total number of model updates, while the total number of samples impacts the diversity of training batches.

Overall, we observe that having more training samples, even at fixed number of updates and thus computing time, drastically improves the accuracy. Specifically, both the average error and the variance decrease with increasing sample size, down to a variance of about 0.1 on $n_\sigma$ in our tests.  Note that this residual variance includes errors from the NF models as well as the uncertainty from the Monte-Carlo estimate (required to be smaller than 0.05).
As expected, higher levels of tensions require more training samples, although with a relatively mild dependence on dimension, confirming the NF method to be fairly robust to dimensionality. Overall, the NF method, when provided with sufficiently many samples, performs extremely well, with biases below $0.2\sigma$, even for a $4\sigma$ tension.
However, we also find that, when using only 1000 training samples, MAF models tend to overfit (as can be observed by comparing the training and validation losses), even though the same models, trained on the same number of batches, perform well on larger training samples. In these cases, the method often is unable to produce a shift probability using the Monte-Carlo sampling technique, except in very low dimension. This occurs when all samples drawn from the approximate posterior $q$ fall within the $q(\vb{0})$ iso-hypersurface, which is itself poorly determined due to overfitting.
In the next section, we provide some practical tools to monitor training and mitigate for under- and overfitting.

\subsection{Strong non-Gaussianity} \label{Sec:NonGaussianExample}

\begin{figure}
\centering
\includegraphics{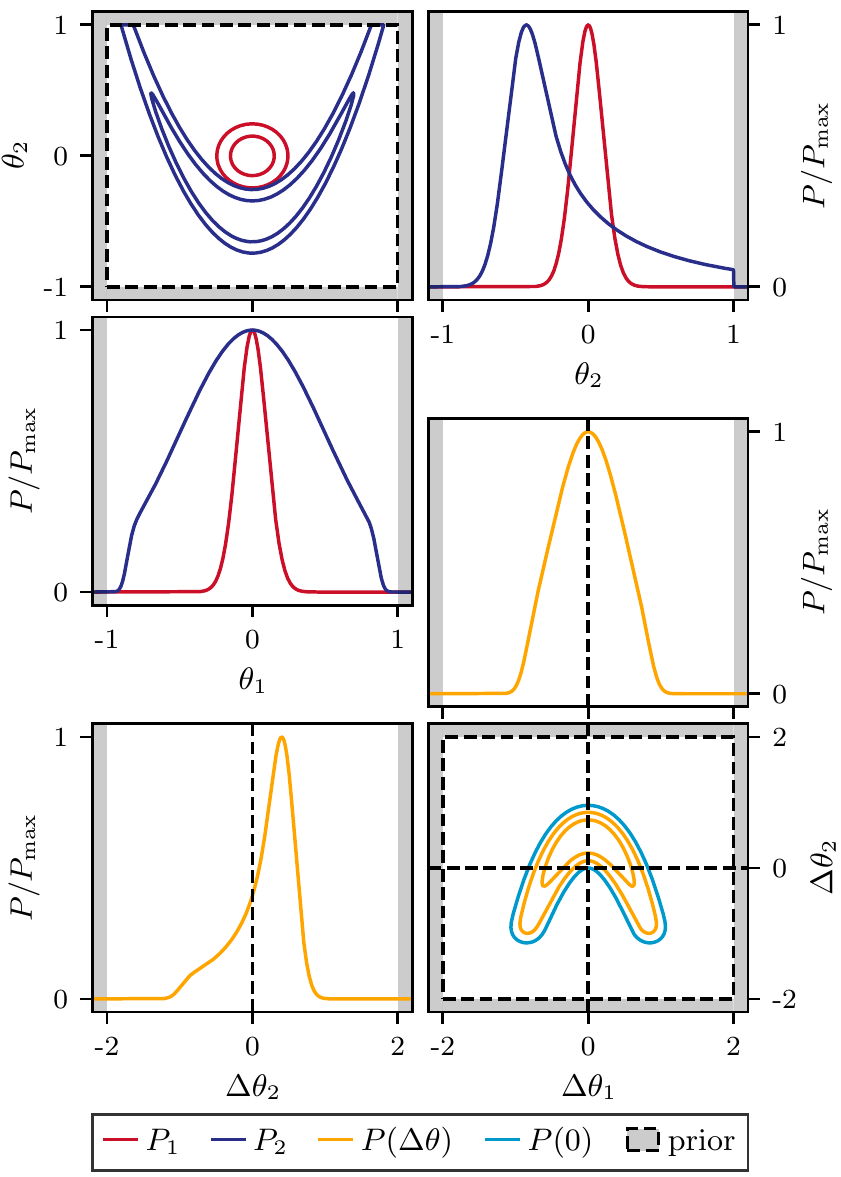}\\
\caption{\label{Fig:NonGaussianExample}
The relevant posterior distributions of the first benchmark example in \cref{Sec:NonGaussianExample}.
The upper triangle plot shows the posteriors for the two single datasets while the lower triangle shows the posterior of parameter differences.
We highlight both the prior boundary and the iso-contour that is compatible with zero parameter shift.
}
\end{figure}

We now consider a first example where one of the two distributions has a strong, banana-shaped, non-Gaussian degeneracy.
We show in \cref{Fig:NonGaussianExample} the two distributions that we use and notice that the strong non-Gaussianity would not be clearly visible in one dimensional projections.
The parameter difference distribution, shown in \cref{Fig:NonGaussianExample}, inherits the strong non-Gaussianity that causes the input tension not to appear appreciably in one dimensional marginalized posteriors.

In this example the first distribution is a multivariate Gaussian with zero mean and diagonal covariance equal to $\sigma^2 = 0.01$ while the second distribution is given by $P_2 \sim \mathcal{N}(\sqrt{\theta_1^2 + 20(2 \theta_1^2-\theta_2-1/2)^2}, 1/4)$, up to an irrelevant normalization constant.
The prior is flat for both parameters in the interval $[-1, 1]$.

The exact input tension for this example is $2.77\sigma$ as computed on a very fine grid that makes the error on this determination irrelevant.
The Gaussian parameter shift $Q_{\rm DM}$ estimator applied to the two datasets reports a tension of $0.12\sigma$ which is largely underestimated. This happens because the long tails of the second distribution move the Gaussian approximation toward the first distribution.
From the two single distributions we can compute the joint distribution as the product of the two. We can then use this joint distribution to compute the significance of update parameter shift, $Q_{\rm UDM}$, applied to the most Gaussian of the two distributions and the joint distribution. This leads to an estimate of $2.81\sigma$ tension which is remarkably accurate. The reason for this is that the effect of the non-Gaussian tails is mitigated in the joint distribution. 


\begin{figure}
\centering
\includegraphics{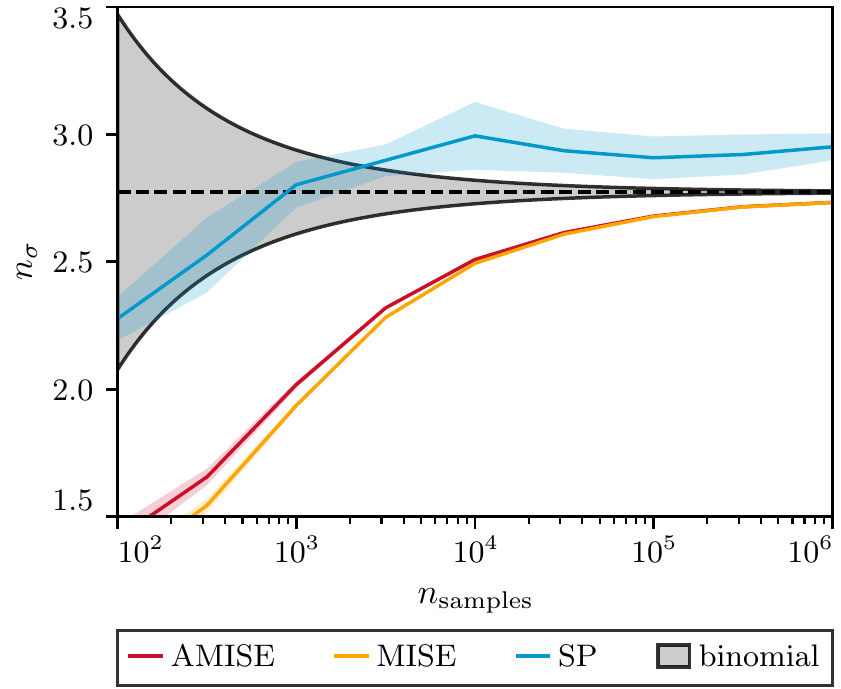}\\
\hspace{0.5cm}
\includegraphics{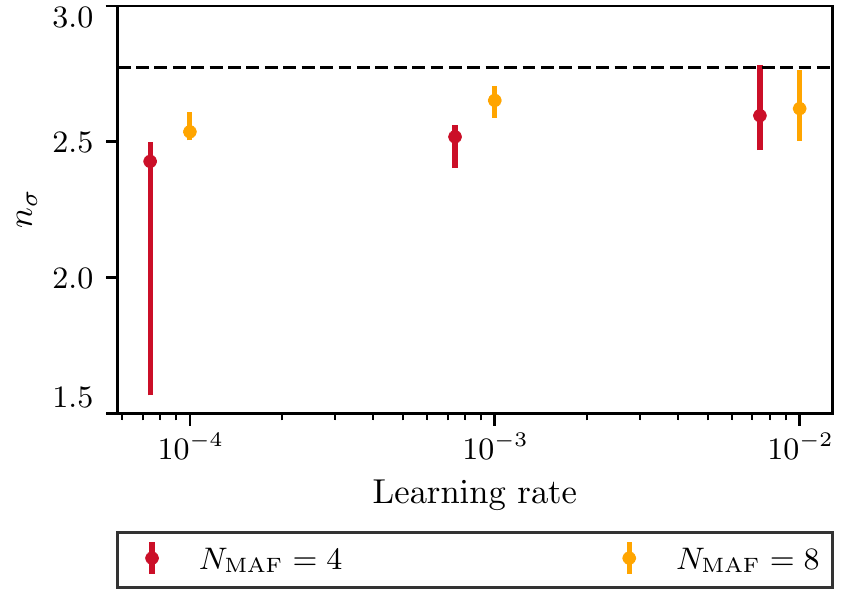}\\
\caption{\label{Fig:NonGaussianExampleKDE}
Determination of the tension in the non-Gaussian example in \cref{Sec:NonGaussianExample} as a function of number of samples for KDE estimates in the upper panel and as a function of learning rate for NF estimates in the lower panel.
The exact result is a $2.77\sigma$ tension as shown by the black dashed line.
In the upper panel the continuous black lines shows the limiting case of binomial scaling of errors.
}
\end{figure}

In the upper panel of \cref{Fig:NonGaussianExampleKDE} we show the scaling of KDE estimates with number of samples.
As we can see the MISE/AMISE estimates increasingly agree as the number of samples increases and show a convergent behavior.
While the power law scaling of the MISE estimate follows the binomial one, $\Delta n_\sigma \propto n^{-1/2}$, it does suffer from a factor ten penalty.
As we can clearly see in \cref{Fig:NonGaussianExampleKDE} these two estimators always have a systematic bias that is larger than their variance.
Hence the best computation strategy is to use the largest sample size available.

On the other hand the SP estimate is extremely precise in the regime where the number of samples is very limited and samples are sparse. In this regime the accuracy of the SP estimator scales like the binomial but with better overall performances. Once the sample size exceeds $10^3$ and the sparsity of samples decreases, the estimator shows no further improvement. In this regime, we notice that the variance of the SP estimator is of the same order of magnitude as its bias. This suggests that, when a large sample is available, one can increase the precision of the estimator by computing it on smaller batches and averaging.
The lack of asymptotic improvement at increasing sample size, for the SP estimator, is due to the absence of a tuning parameter, the smoothing scale, that in the MISE/AMISE case changes as sample size increases.

\begin{figure}
\centering
\includegraphics{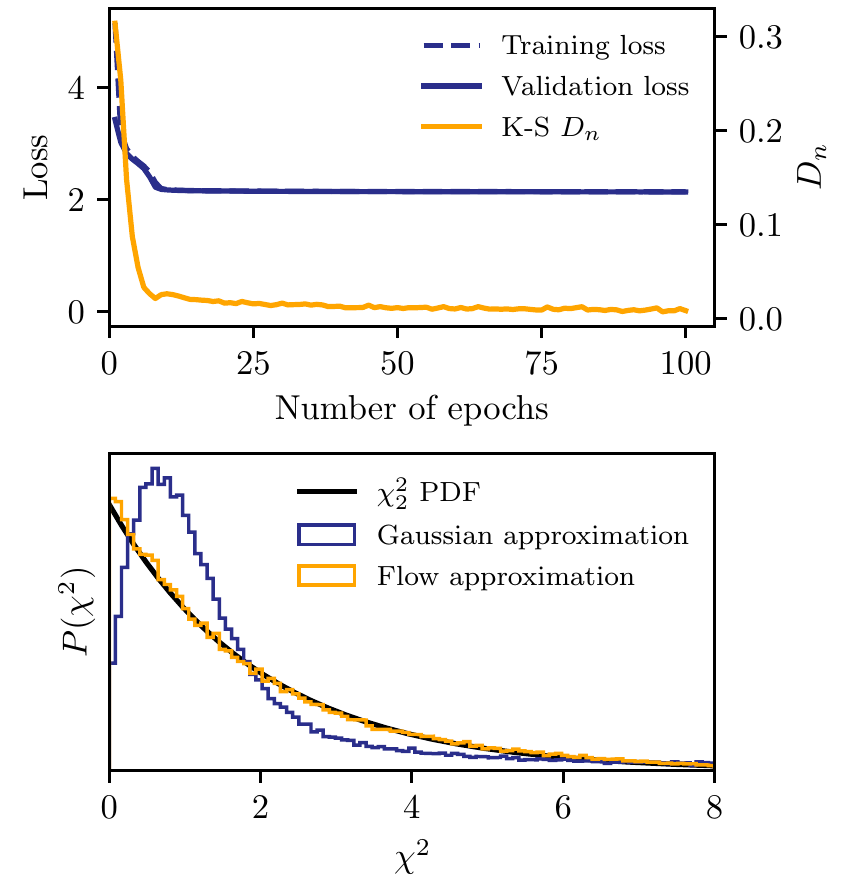}\\
\includegraphics{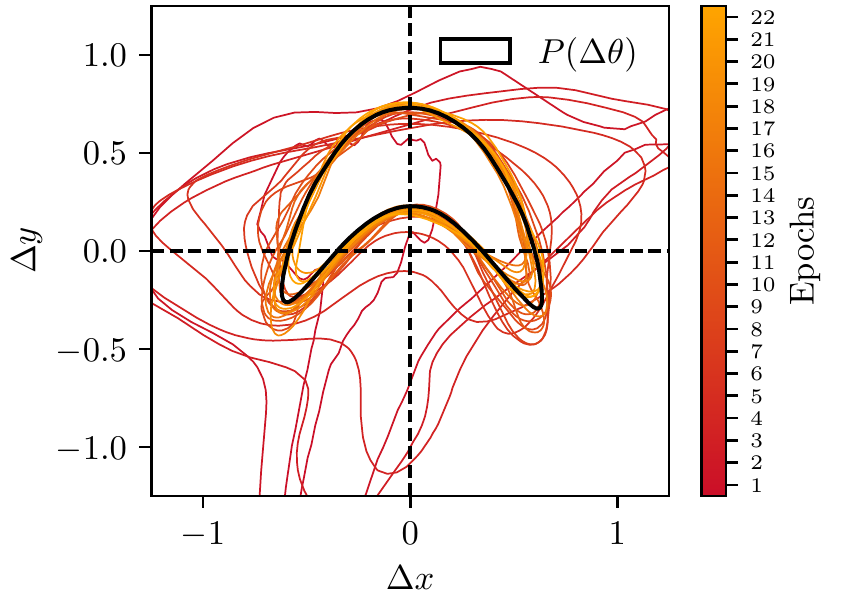}\\
\caption{\label{Fig:ExTraining}
Illustration of training diagnostics for the NF method. The upper panel shows the evolution of training {(solid)} and validation {(dashed)} losses, in blue, as a function of the number of epochs {(the two almost completely overlap)}. At the end of every epoch, we compute the squared norms of the transformed samples and perform a Kolmogorov-Smirnov (K-S) test, for which the null hypothesis is that these norms are $\chi^2_d$-distributed (here $d=2$). We show the evolution of the K-S statistic $D_n${, in yellow,} in the upper panel on a different y-axis. In the middle panel, we show the distributions of squared norms of transformed samples, first through the Gaussian approximation (blue) and through the trained MAF (yellow). In practice, we draw these plots at the end of every epoch, allowing the user to monitor convergence. The lower panel illustrates the learning the difference distribution, by showing the 95\% contours of the approximate distribution {at the end of} the first {twenty} epochs of training (colored contours). The true distribution is shown in black.
}
\end{figure}

In the lower panel of \cref{Fig:NonGaussianExampleKDE} we show the error of NF estimates as a function of learning rate. Note that here, we maintain the value of the learning rate fixed, which determines the sensitivity to local optima, while for realistic cases, we recommend decreasing it progressively during training (see \cref{Sec:DESvsPlanck}). We observe that if the learning rate is set too low, the NF model tends to underfit the banana-shaped distribution, leading to an underestimation of the tension. However, setting the learning rate at $10^{-2}$, the maximum value we used in our test, increases the variance in the estimate, as can be seen by the wider boxes in \cref{Fig:NonGaussianExampleKDE}. We also note that in this example, the more flexible models with eight stacked MADEs generally show less variance than models with only four. A possible reason is that MADEs, which implement auto-regressive operations, are sensitive to the ordering of variables, which is why we randomly permute them between MADEs. Therefore, since the difference distribution shows strong non-Gaussianity in only one of the two dimensions, models with more MADEs are likely to operate on the most efficient orderings of variables, at the price of some redundancy. 

Finally, we illustrate training diagnostics in \cref{Fig:ExTraining}. Before training, the sample is randomly split into a training sample and a smaller validation sample, which is not used for training, to watch for overfitting. The upper panel shows the evolution, as a function of training epochs, of the training and validation losses (given by the average log-probability of the tranformed samples, see \cref{Sec:NFmethod}) in blue as well as the evolution of the Kolmogorov-Smirnov statistics, $D_n$, {in yellow. This quantity is} testing that the squared norms of samples in the gaussianized space follows a $\chi^2_2$ distribution. This gaussianization can be visualized in the {middle} panel of \cref{Fig:ExTraining} where we compare the distribution of the squared norms of samples in the Gaussian approximation space (blue histogram) and in the fully gaussianized space (yellow histogram) to the $\chi^2_2$ distribution. {Finally, we illustrate in the lower panel of \cref{Fig:ExTraining} how the model evolves during training and approaches the true distribution (black), by showing the learned distribution at the end of the twenty first epochs (red to yellow).}

\subsection{Informative prior} \label{Sec:InformativePriorExample}

\begin{figure}
\centering
\includegraphics[width=\columnwidth]{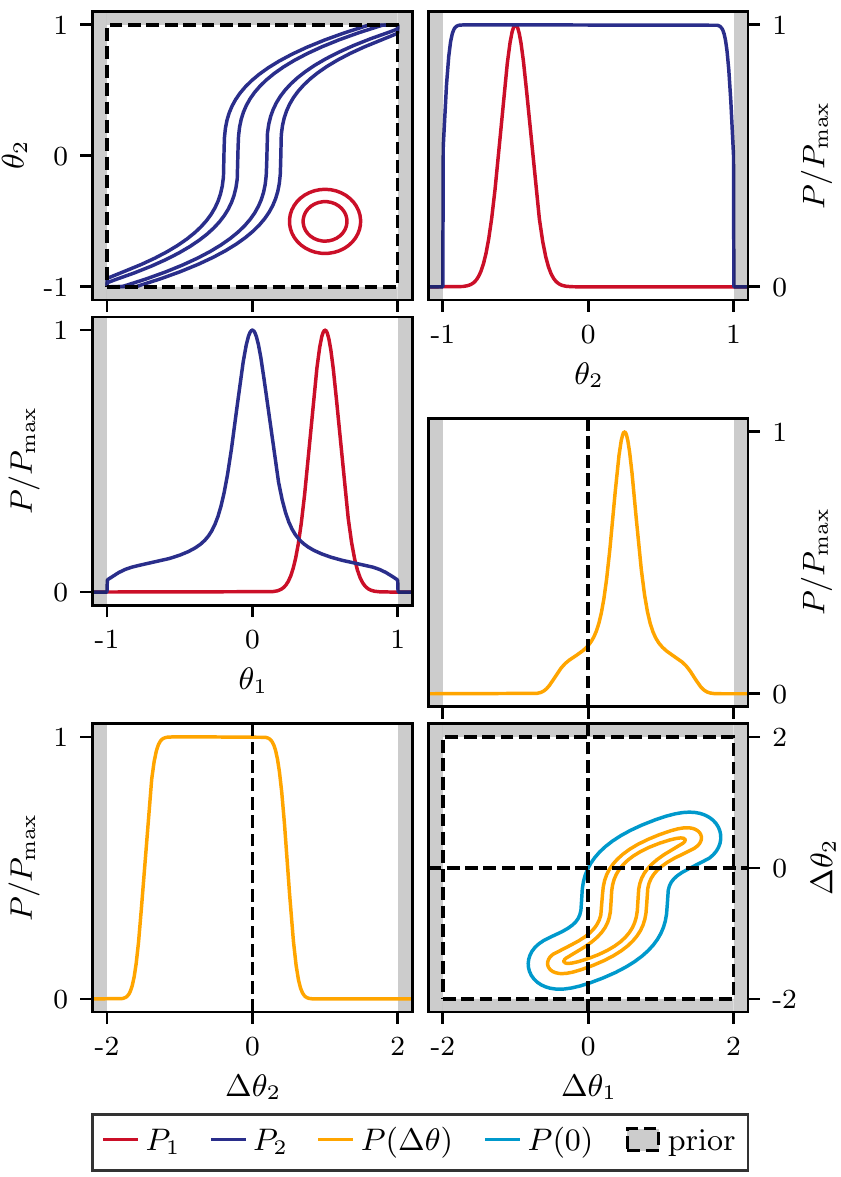}\\
\caption{\label{Fig:InfoPriorExample}
The relevant posterior distributions of the second benchmark example in \cref{Sec:InformativePriorExample}.
The upper triangle plot shows the posteriors for the two single datasets while the lower triangle shows the posterior of parameter differences.
We highlight both the prior boundary and the iso-contour that is compatible with zero parameter shift.
}
\end{figure}

We now consider an example in which the prior is informative.
We show in \cref{Fig:InfoPriorExample} the two distributions that we use.
This example has the highest tension we consider, yet that can hardly be guessed from one dimensional marginalized posteriors.
In this example the first distribution is a multivariate Gaussian with mean $(1/2, -1/2)$ and diagonal covariance equal to $\sigma^2 = 0.01$, while the second distribution is given by $P_2 \sim \mathcal{N}(x-y^3, 1/2)$.
The second distribution has a direction which is fully degenerate and constrained by the prior on the unity square.
\Cref{Fig:InfoPriorExample} also clearly shows that, while the prior boundary introduces a discontinuity in the two posterior distributions it does not do so for the parameter difference distribution.

The exact input tension for this example is $4.03\sigma$ computed as in the previous example.
The Gaussian estimator applied to the two datasets reports a tension of $3.93\sigma$ while the Gaussian update estimator reports $3.89\sigma$.
Both Gaussian estimates agree fairly well with the exact result mostly because the tension in the correlated direction is so high that details of the degeneracy or the prior hardly change the end result.


\begin{figure}
\centering
\includegraphics{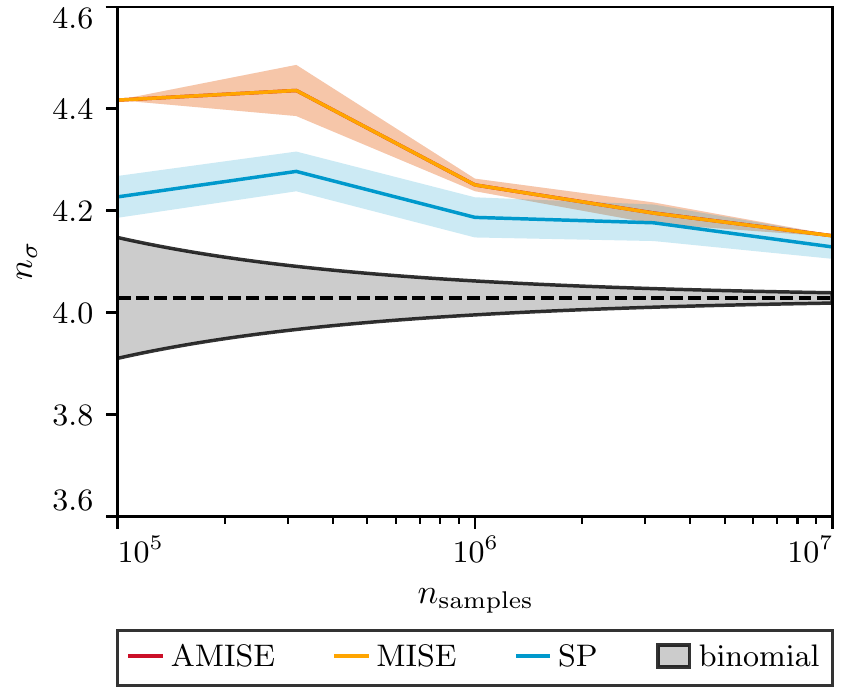}\\
\hspace{0.5cm}
\includegraphics{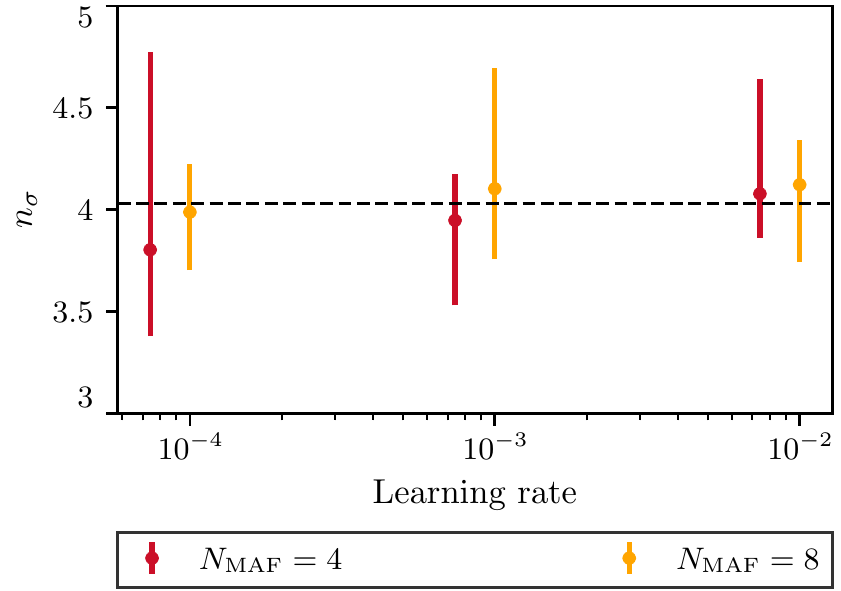}\\
\caption{\label{Fig:InfoPriorExampleKDE}
Determination of the tension in the informative prior example in \cref{Sec:InformativePriorExample} as a function of number of samples for KDE estimates in the upper panel and as a function of learning rate for NF estimates in the lower panel.
The exact result is a $4.03\sigma$ tension as shown by the black dashed line.
In the upper panel the continuous black lines shows the limiting case of binomial scaling of errors.
}
\end{figure}

In \cref{Fig:InfoPriorExampleKDE} we show the absolute error scaling of KDE estimates with number of samples.
As we can see a meaningful determination of the result requires significantly more samples with respect to the previous examples, compatibly with the input tension which is sensibly larger.

The number of samples is so large that the MISE and AMISE estimates entirely agree.
In this example the tension is large so the SP estimator is expected to have good performances as can be clearly seen in \cref{Fig:InfoPriorExampleKDE}. In particular samples are so sparse around the zero iso-contour, due to the high tension, that the SP estimator has similar performances to the MISE one with an order of magnitude less samples.


NF estimates show good performance, somewhat independent of the learning rate. However, NFs with eight stacked MADEs generally show lower errors, especially at lower learning rate, because of the more complex morphology of the difference distribution. The best case is found for MAFs made of eight MADEs and the smallest learning rate, in contrast with the previous case, prone to overfitting. In this case, we find errors of order 0.2 with negligible bias (over ten runs).
We note that the spread is larger for this example than the previous one. Indeed, we observed more important fluctuations of the predicted iso-contour of zero shift over the ten runs, consistent with the fact that training samples are sparser in this case where the true tension if higher.

\subsection{Multi-modal posterior} \label{Sec:MultiModalExample}

\begin{figure}
\centering
\includegraphics[width=\columnwidth]{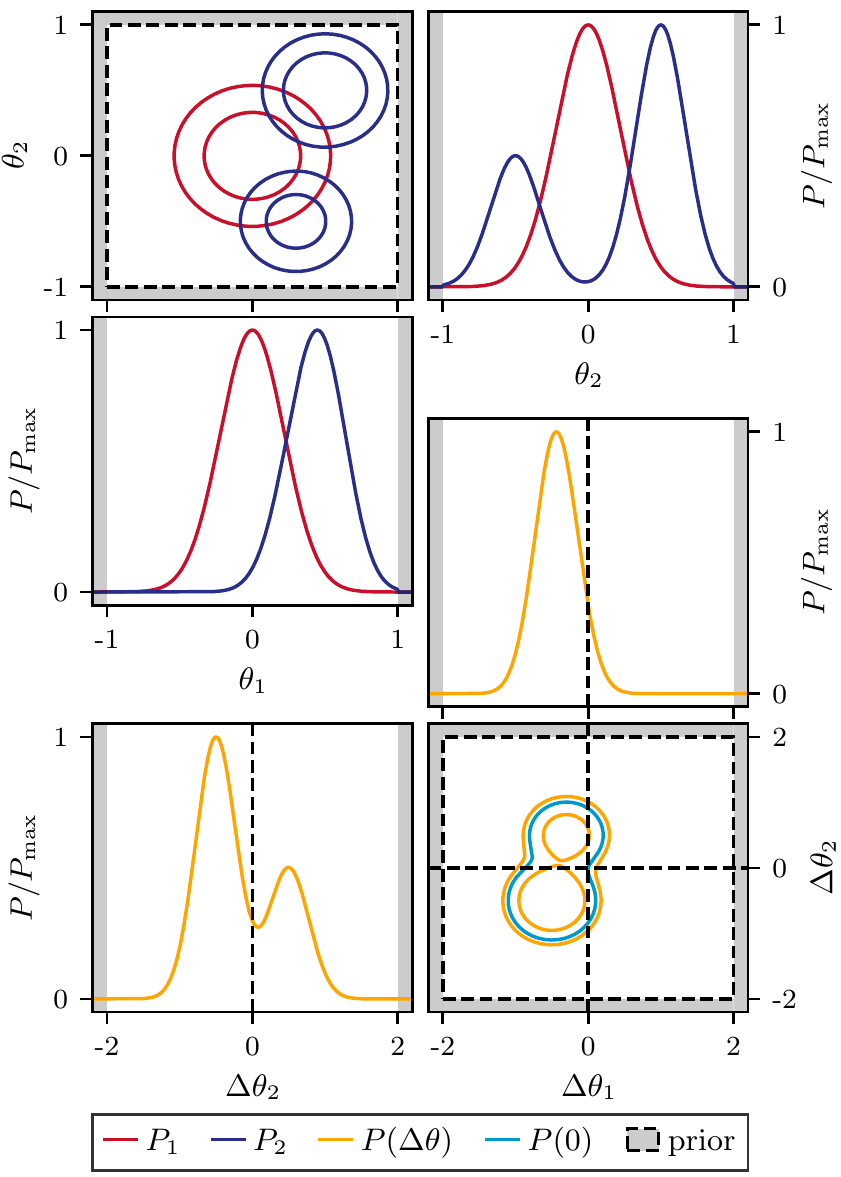}\\
\caption{\label{Fig:MultiModExample}
The relevant posterior distributions of the third benchmark example in \cref{Sec:MultiModalExample}.
The upper triangle plot shows the posteriors for the two single datasets while the lower triangle shows the posterior of parameter differences.
We highlight both the prior boundary and the iso-contour that is compatible with zero parameter shift.
}
\end{figure}

We last consider an example that entails a multi-modal posterior.
We show in \cref{Fig:MultiModExample} the two distributions that we use.
The first distribution is a multivariate Gaussian with zero mean and diagonal covariance equal to $\sigma^2 = 0.05$ while the second distribution is a mixture of two Gaussian distributions, with weights $(2,1)$, means $(0.5, 0.5)$ and $(0.3, -0.5)$ and diagonal covariances both equal to $\sigma^2 = 0.03$.

The exact input tension for this example is $1.66\sigma$ computed as in the previous examples on a grid.
The Gaussian estimator applied to the two datasets reports a tension of $1.07\sigma$ while the Gaussian update estimator gets closer to the correct result and reports a tension of $1.40\sigma$.


\begin{figure}
\centering
\includegraphics{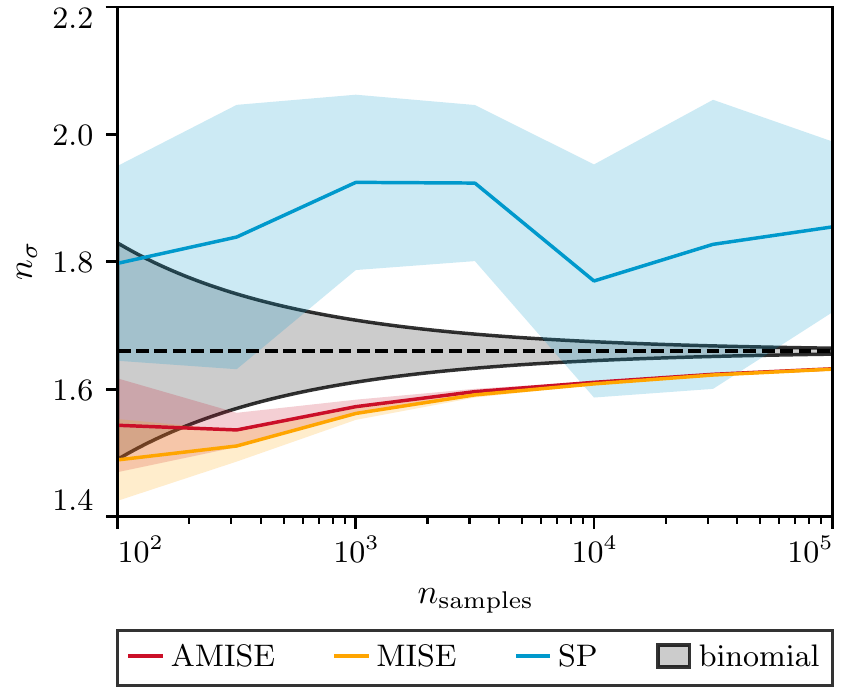}\\
\hspace{0.5cm}
\includegraphics{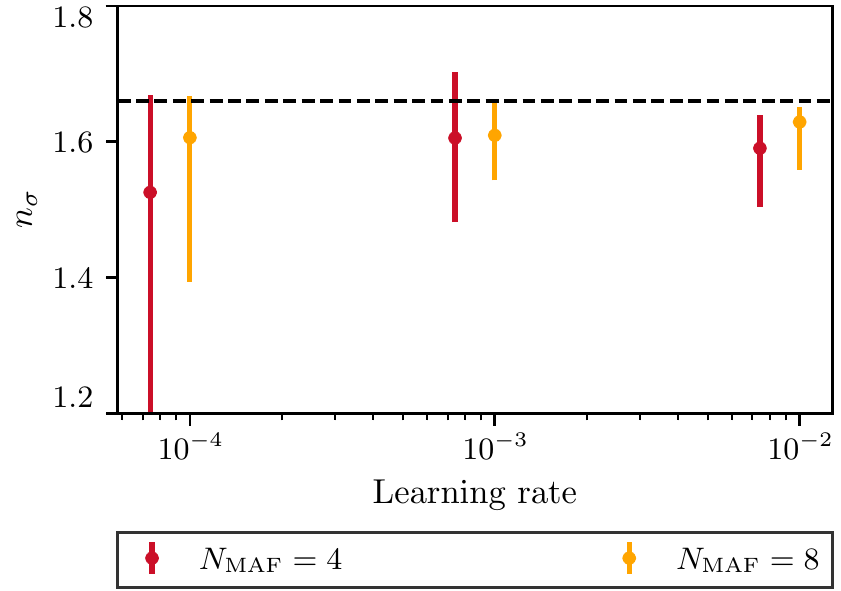}\\
\caption{\label{Fig:MultiModExampleKDE}
Determination of the tension in the multi-modal example in \cref{Sec:MultiModalExample} as a function of number of samples for KDE estimates in the upper panel and as a function of learning rate for NF estimates in the lower panel.
The exact result is a $1.66\sigma$ tension as shown by the black dashed line.
In the upper panel the continuous black lines shows the limiting case of binomial scaling of errors.
}
\end{figure}

\Cref{Fig:MultiModExampleKDE} shows the behavior of the KDE estimators that mostly follows previous examples.
The large number of samples and the relatively low tension result in fairly poor performances of the SP estimator while the MISE/AMISE estimators show an absolute error decay that is slower than the binomial power law.

Multi-modal distributions can be challenging to model with normalizing flows, especially with simple autoregressive operations. However, we observe good performance of our NF method in this example, as shown by the small errors in \cref{Fig:MultiModExampleKDE}, albeit with slightly underestimated results. We find that using larger learning rates provides more accurate results, similar to the first example, and models with eight MADEs show lower variance.

\section{Application to DES~Y1 and Planck} \label{Sec:RealExample}

In the previous section we have thoroughly investigated the behavior of the two techniques to compute the significance of parameter differences with benchmark examples where the answer is known.

In this section we apply these estimators to real data where the correct answer is not known, showing a use case with realistic challenges.
We consider two datasets: the clustering and lensing of galaxies from the Dark Energy Survey (DES) Year 1 dataset~\cite{Abbott:2017wau}; the Planck 2018 measurements of CMB temperature and polarization at small (Planck 18 TTTEEE) and large angular scales (lowl+lowE)~\citep{Aghanim:2018eyx, Aghanim:2019ame}.

With these two data sets we investigate two cases. 
In \cref{Sec:DESInternal}, we quantify the internal consistency of DES cosmic shear measurements with the DES measurements of the galaxy-galaxy lensing and galaxy clustering two-point functions. 
This case shows appreciable data correlation.
In \cref{Sec:DESvsPlanck}, we compute the tension between DES Y1 and Planck 2018, as an example that does not have appreciable correlation between the two data sets.

\subsection{DES~Y1 cosmic shear \vs galaxy-galaxy lensing and clustering} \label{Sec:DESInternal}

\begin{figure}
\centering
\includegraphics{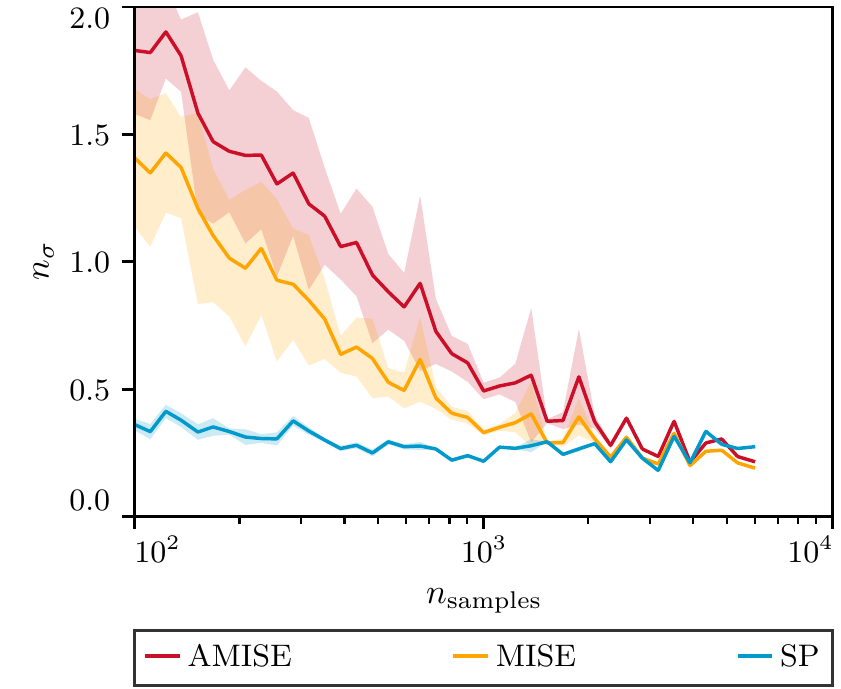}\\
\caption{\label{Fig:DESInternal}
Internal tension between DES~Y1 lensing and galaxy clustering, including their cross correlation, as in \cref{Sec:DESInternal}, as a function of number of samples, according to KDE estimates.
Different lines show different algorithms, as shown in legend.
}
\end{figure}

\begin{figure}
\centering
\includegraphics{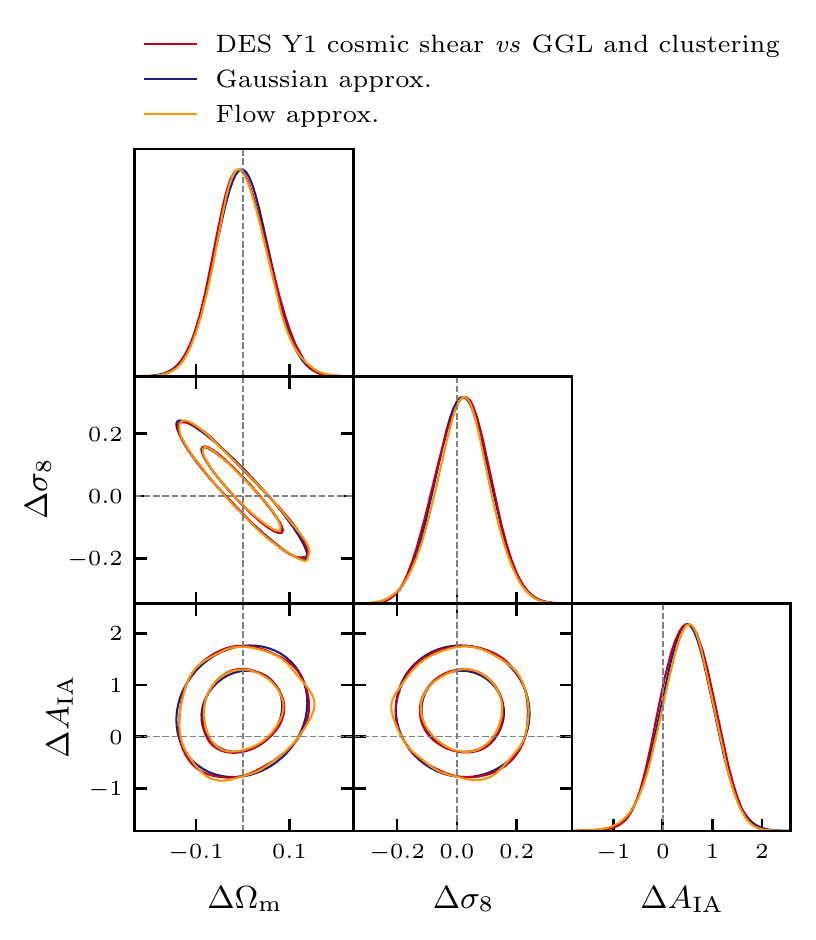}\\
\caption{\label{Fig:DESInternalCorner}
Marginalized parameter difference distribution between DES~Y1 cosmic shear and the combination of galaxy-galaxy lensing (GGL) and clustering, including their cross correlation. We only show the subspace of the most relevant parameters.
}
\end{figure}

In this section, we investigate the agreement between cosmic shear measurements from DES Y1 with those of galaxy-galaxy lensing and galaxy clustering.

These data sets are strongly correlated, which means that we must apply the parameter duplication scheme detailed in \cref{Sec:ParamDifferences} to sample the parameter difference posterior. Here, we duplicate the six cosmological parameters used in the DES~Y1 analysis, including the sum of neutrino masses, as well as the two parameters describing intrinsic alignments (amplitude and redshift dependence). We do not duplicate galaxy clustering biases as they only impact the galaxy-galaxy lensing and clustering measurements, and we opt not to duplicate other nuisance parameters that are strongly prior dominated and would considerably increase the computing time needed to sample the parameter difference posterior.
To sample the parameter difference distribution, we slightly modify the DES Y1 analysis pipeline within \cosmosis~\cite{Zuntz:2014csq} to compute the theoretical predictions for the two subsets of the data vector with their respective copies of the parameters, and then combine them in the full likelihood. We sample the posterior with \multinest~\cite{Feroz:2007kg} and show 68\% and 95\% contours in \cref{Fig:DESInternalCorner} in red.
We find that the maximum correlation coefficient between all parameters is $58\%$ (see App~A of \cite{Raveri:2019gdp}), meaning that it is indeed crucial to account for the correlation between data sets.
We note that this analysis in parameter space is complementary to the extensive analysis performed in data space in \cite{Doux:2020kdz}.

We start by computing Gaussian estimators and find that standard parameter difference yields a tension of $n_\sigma = 0.08$ while its update form gives $n_\sigma = 0.5$. In this case we apply the Gaussian estimators to the duplicate parameters to account for the data correlation, as discussed in~\cite{Raveri:2019gdp}.

In this example the number of samples in the parameter difference chain is limited by the number of samples in the nested sampling chain and it is sensibly lower than what can be obtained in case of uncorrelated samples.
For this reason we vary the sample pool for the KDE estimation from $10^2$ samples to all samples in the chain.
The number of repetitions of the calculation is then limited by the random non-overlapping samples that we can draw.
As we can see, similarly to the previous case, for very small sample sizes, all KDE estimates are largely spread and converge toward $n_\sigma = 0.4$ as sample size increases.
In this case we notice that the bias of the SP estimator is very low even at small sample size.

For NF, we also employ the default MAF architecture with 16 MADEs, for a total of 11008~trainable weights. With 10 runs, we find $n_\sigma=0.23\pm0.02$, with a small variance that can be attributed to the fact that the difference distribution is very close to Gaussian in this case, as can be seen on \cref{Fig:DESInternalCorner}. This result is consistent with the results obtained with KDE using the full sample.
This test shows good agreement between the subsets of the DES Y1 data used here, consistent with~\cite{Doux:2020kdz}. Using a comparison in data space (asymmetric in the two data sets), they found a calibrated $p$-value of 0.396 for the ``cosmic shear \vs galaxy-galaxy lensing and clustering'' consistency test, which may be compared to our probability of no shift of $\Delta=0.29$. 

\subsection{DES~Y1 3x2pt \vs Planck} \label{Sec:DESvsPlanck}

\begin{figure}
\centering
\includegraphics{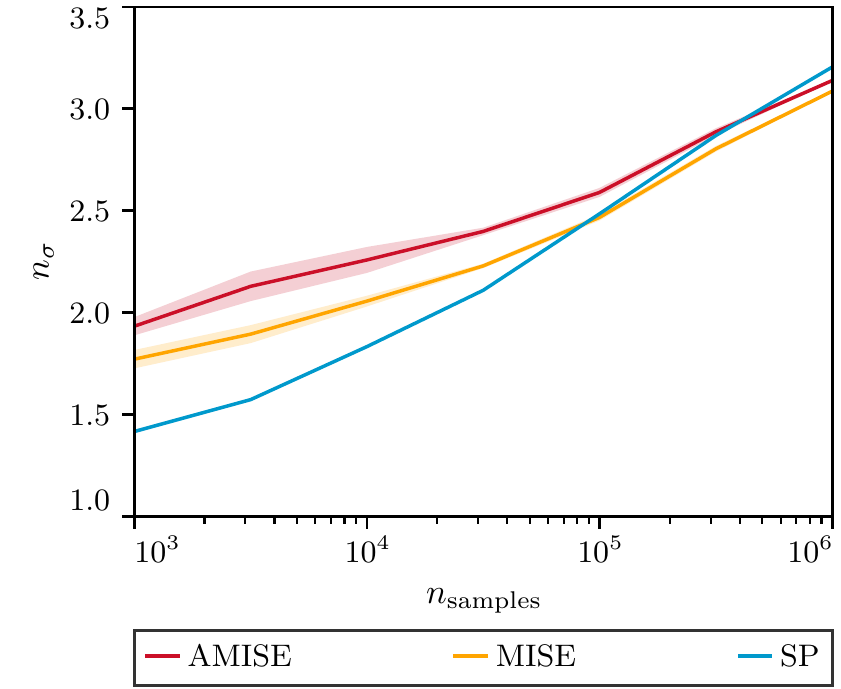}\\
\caption{\label{Fig:DESvsPlanck}
Tension between DES and Planck as in \cref{Sec:DESvsPlanck} as a function of number of samples, as estimated through the KDE algorithm.
Different lines show different algorithms, as shown in legend.
}
\end{figure}

For this example, we start by sampling the posterior for each data set, varying the six parameters of the $\Lambda$CDM model with the non-informative priors used in the Planck 2018 analysis~\cite{Aghanim:2019ame}. We also include all the recommended parameters and priors describing systematic effects in the different data sets we consider. 
In this case, the sum of neutrino masses is fixed to the minimal value of {0.06~eV}~(e.g.~\cite{Long:2017dru}).

We start by computing the Gaussian tension between the two data sets we consider.
As we have seen in the previous section, the standard Gaussian formula generally misestimates tensions and in this case reports a tension of $n_{\sigma} = 0.5$.
The update Gaussian parameter shift on the other hand, in presence of at least one Gaussian data set (Planck), is expected to be precise. In this case the UDM estimator reports $n_{\sigma} = 3.0$, which we can use to gauge the order of magnitude of the tension.
As such this allows us to orientate algorithmic choices like the number of samples that we would need.
The UDM estimator also reports that only two of the six parameters we are considering are significantly contributing to the cosmological parameter constraints and to this tension determination while the other are prior dominated.
The data constrained parameters are combinations of $\sigma_8$ and $\Omega_m$ that are given by
$\sigma_8 \Omega_m^{0.7}$ and $\Omega_m / \sigma_8$.
Even though prior dominated parameter combinations should not contribute to the end result, as we have shown in \cref{Sec:FlatPosterior}, we still include them in this calculation.

From the binomial distribution, we would expect to need at least $5 \times 10^3$ samples to quantify a $3$ sigma tension with an accuracy of $\Delta n_\sigma = 0.1$. We then increase the number of samples from $10^3$ to $10^6$ and compute the KDE estimator.
\Cref{Fig:DESvsPlanck} shows the behavior of the KDE estimator at increasing number of samples, averaging the result over ten pools of samples randomly drawn.
As we can see, at low number of samples, all KDE estimators are largely biased with respect to their end result, and with a significant spread of about $0.5\sigma$.
As the number of samples increases the spread between the different estimates reduces and the different estimators converge to about $n_\sigma = 3.1$.
This slow convergence is due to the large number of prior constrained directions that are hard for the KDE estimator to resolve accurately.
We verified that, using only the parameters that are actually involved in the physical origin of this tension,
the result would give a determination of $n_\sigma = 3.1$ with small variation across sample size.

\begin{figure*}
\centering
\includegraphics{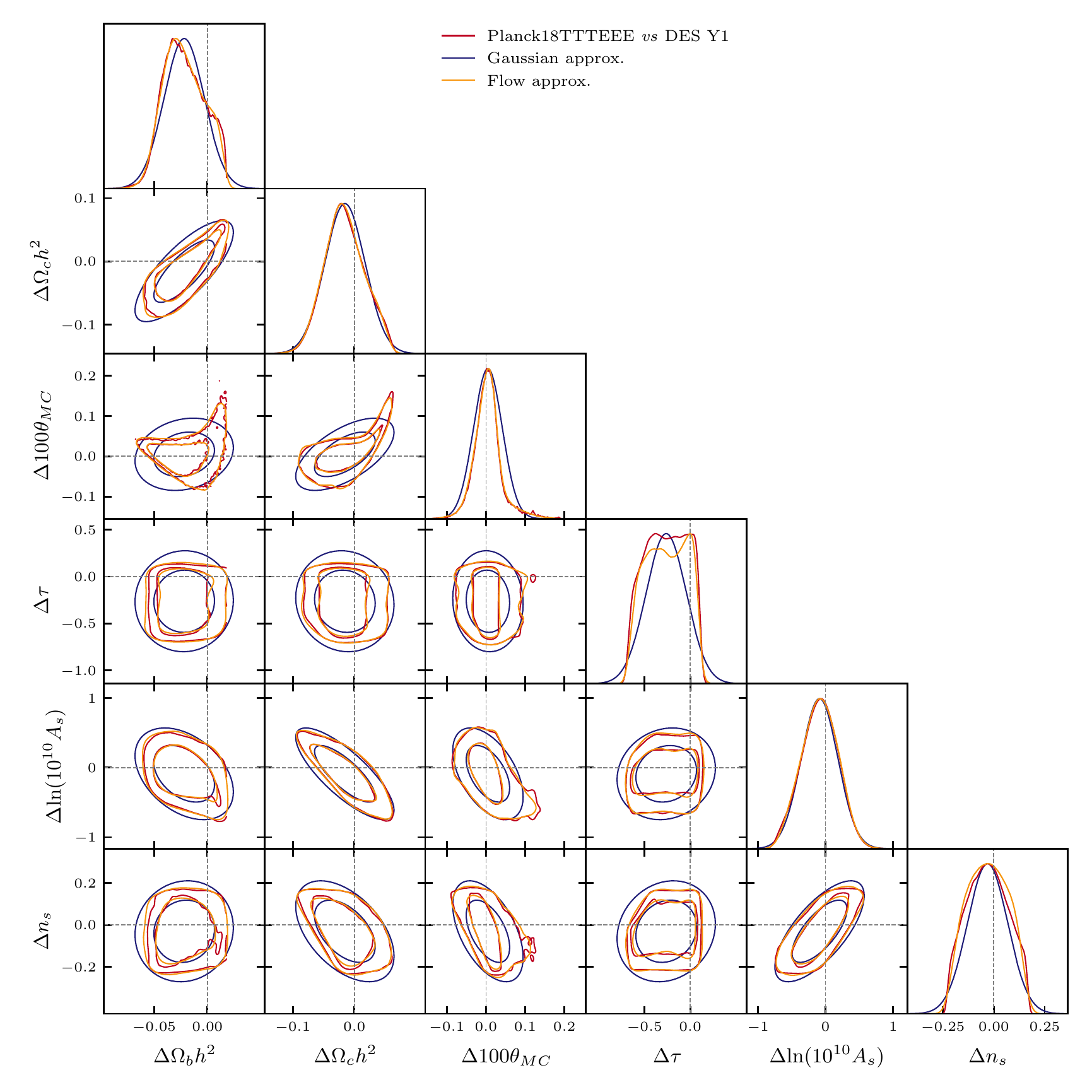}
\caption{ \label{fig:my_label}
Parameter difference distribution between DES~Y1 and Planck 18 TTTEEE. The true distribution is shown in red, its Gaussian approximation in blue and the normalizing flow approximation in yellow. The NF approximation captures non-Gaussian features of the distribution, including those due to the prior (e.g. on the $\tau$ parameter).}
\end{figure*}

For NF, we employ the default MAF architecture suggested in \cref{Sec:FlowEstimate} with 12 MADEs, for a total of 4752~trainable weights. Following guide lines drawn from benchmark examples, we initialize the learning rate at a sufficiently high value of $10^{-2}$ and decrease it during training using the \texttt{ReduceLROnPlateau} callback implemented in \texttt{tensorflow.keras.callbacks}. We perform 10 runs to assert the level of stochasticity of the estimator and find a average tension of $\expval{n_\sigma}=3.0$ with a standard deviation of 0.12 (with extrema at 2.78 and 3.13). We conclude that the suggested architecture provides sufficiently accurate estimates of tension in this real-data example. Finally, in order to demonstrate the quality of the approximation $q$ to the parameter difference distribution $\Post$, we overlay in \cref{fig:my_label} the 68 and 95\% contours for the real chain samples (in red) and for a sample of $10^4$ points drawn from the approximation $q$. The agreement between the two sets of contours shows that the NF model can learn non-Gaussian features of the distribution, including those induced by the complex prior (e.g. for the $\tau$ parameter).

\section{Conclusions} \label{Sec:Conclusions}

As the precision of different cosmological data sets increases, so does the need to reliably check and monitor their consistency.
In this paper we have studied a method for doing so that relies on and exploits the full distribution of differences between the posteriors of two experiments.
We have characterized the properties of this {\it parameter difference} estimator showing that it allows to characterize tensions between probes, making no assumption about the underlying distribution of parameters.

We have developed two complementary algorithms for the efficient calculation of the statistical significance of a parameter difference and shown that they generally achieve fully workable performances through a series of toy examples.
These show that a tension ranging from 1$\sigma$ to 4$\sigma$ in dimension 2 to 30 can be accurately calculated with an error that is below a fraction of a sigma.

The first method we develop relies on kernel density estimates (KDE) of the posterior with fixed and variable smoothing scale. We found that estimates with fixed smoothing scale work best in low dimensions and show fast convergence.
Variable smoothing scale estimators, on the other hand, show slower convergence but largely outperformed fixed smoothing scale estimator in high dimensions with limited number of posterior samples.
We have developed an algorithm that solves the problem of the KDE estimator that would naively scale as the number of posterior samples squared, while our {\it neighbor elimination} algorithm scales as $\order{n\log n}$, outperforming FFT based methods in dimension larger than two.

The second method we develop uses normalizing flows to learn and model the parameter difference posterior distribution from its samples and provides a first use case for machine learning techniques at the problem of estimating tensions between data sets.
This technique shows good scaling to large number of samples and high dimension, benefiting from efficient implementations of normalizing flows algorithms in \textsc{TensorFlow-Probability}. In particular, the running time for the normalizing flow estimate is determined by the number of updates during training rather than the number of samples. The more samples are available, the more diverse mini-batches are and the faster the model converges.

The two calculation methods we presented have to agree exactly on the statistical significance of a tension.
Since they are profoundly different in the way statistical significance is obtained, their spread can be used as a reliable global error estimate, as both results can be obtained in reasonable time on a laptop (from few minutes to few hours). Our benchmark experiments show that the two methods agree with each other and the true tension within $0.5\sigma$ {in difficult cases, and generally within $0.1-0.2\sigma$} or better, provided enough samples are used.

We have showcased their use with two examples from cosmology.
In the first one we have quantified the internal agreement at parameter level of galaxy clustering and lensing from the first year release of DES, finding them in excellent agreement, at the $0.3\sigma$ level.
We have then quantified the level of agreement between CMB power spectrum measurements from Planck and DES finding them in a significant tension equivalent to $3 \sigma$.
This is slightly larger than what reported in~\cite{Lemos:2020jry} since we fix neutrino masses to their minimal value, thus increasing the precision of other parameter determinations and hence their tension.

Looking forward, several improvements to the techniques we discuss would be worth pursuing. 
For KDE methods these would include the development of smoothing scale selectors optimized at the specific iso-contour problem at hand rather than general purpose performances across all parameter space.
For normalizing flow estimates, an improvement will be to simultaneously train an ensemble of models and average the results, with weights based on the quality of the fit of each model.

\acknowledgments

We thank 
Bhuvnesh Jain, Mike Jarvis, Pablo Lemos, Youngsoo Park, Vivian Poulin
for helpful comments.
MR and CD are supported in part by NASA ATP Grant No. NNH17ZDA001N, and by funds provided by the Center for Particle Cosmology.
CD is supported in part by the US Department of Energy Grant No. de-sc0007901.
Computing resources are in part provided by the University of Chicago Research Computing Center through the Kavli Institute for Cosmological Physics at the University of Chicago. 

\vfill
\bibliography{biblio}

\begin{thebibliography}{48}%
\makeatletter
\providecommand \@ifxundefined [1]{%
 \@ifx{#1\undefined}
}%
\providecommand \@ifnum [1]{%
 \ifnum #1\expandafter \@firstoftwo
 \else \expandafter \@secondoftwo
 \fi
}%
\providecommand \@ifx [1]{%
 \ifx #1\expandafter \@firstoftwo
 \else \expandafter \@secondoftwo
 \fi
}%
\providecommand \natexlab [1]{#1}%
\providecommand \enquote  [1]{``#1''}%
\providecommand \bibnamefont  [1]{#1}%
\providecommand \bibfnamefont [1]{#1}%
\providecommand \citenamefont [1]{#1}%
\providecommand \href@noop [0]{\@secondoftwo}%
\providecommand \href [0]{\begingroup \@sanitize@url \@href}%
\providecommand \@href[1]{\@@startlink{#1}\@@href}%
\providecommand \@@href[1]{\endgroup#1\@@endlink}%
\providecommand \@sanitize@url [0]{\catcode `\\12\catcode `\$12\catcode
  `\&12\catcode `\#12\catcode `\^12\catcode `\_12\catcode `\%12\relax}%
\providecommand \@@startlink[1]{}%
\providecommand \@@endlink[0]{}%
\providecommand \url  [0]{\begingroup\@sanitize@url \@url }%
\providecommand \@url [1]{\endgroup\@href {#1}{\urlprefix }}%
\providecommand \urlprefix  [0]{URL }%
\providecommand \Eprint [0]{\href }%
\providecommand \doibase [0]{http://dx.doi.org/}%
\providecommand \selectlanguage [0]{\@gobble}%
\providecommand \bibinfo  [0]{\@secondoftwo}%
\providecommand \bibfield  [0]{\@secondoftwo}%
\providecommand \translation [1]{[#1]}%
\providecommand \BibitemOpen [0]{}%
\providecommand \bibitemStop [0]{}%
\providecommand \bibitemNoStop [0]{.\EOS\space}%
\providecommand \EOS [0]{\spacefactor3000\relax}%
\providecommand \BibitemShut  [1]{\csname bibitem#1\endcsname}%
\let\auto@bib@innerbib\@empty
\bibitem [{\citenamefont {Aghanim}\ \emph
  {et~al.}(2020{\natexlab{a}})\citenamefont {Aghanim} \emph
  {et~al.}}]{Aghanim:2018eyx}%
  \BibitemOpen
  \bibfield  {author} {\bibinfo {author} {\bibfnamefont {N.}~\bibnamefont
  {Aghanim}} \emph {et~al.} (\bibinfo {collaboration} {Planck}),\ }\href
  {\doibase 10.1051/0004-6361/201833910} {\bibfield  {journal} {\bibinfo
  {journal} {Astron. Astrophys.}\ }\textbf {\bibinfo {volume} {641}},\ \bibinfo
  {pages} {A6} (\bibinfo {year} {2020}{\natexlab{a}})},\ \Eprint
  {http://arxiv.org/abs/1807.06209} {arXiv:1807.06209 [astro-ph.CO]}
  \BibitemShut {NoStop}%
\bibitem [{\citenamefont {Aghanim}\ \emph
  {et~al.}(2020{\natexlab{b}})\citenamefont {Aghanim} \emph
  {et~al.}}]{Aghanim:2019ame}%
  \BibitemOpen
  \bibfield  {author} {\bibinfo {author} {\bibfnamefont {N.}~\bibnamefont
  {Aghanim}} \emph {et~al.} (\bibinfo {collaboration} {Planck}),\ }\href
  {\doibase 10.1051/0004-6361/201936386} {\bibfield  {journal} {\bibinfo
  {journal} {Astron. Astrophys.}\ }\textbf {\bibinfo {volume} {641}},\ \bibinfo
  {pages} {A5} (\bibinfo {year} {2020}{\natexlab{b}})},\ \Eprint
  {http://arxiv.org/abs/1907.12875} {arXiv:1907.12875 [astro-ph.CO]}
  \BibitemShut {NoStop}%
\bibitem [{\citenamefont {Riess}\ \emph {et~al.}(2021)\citenamefont {Riess},
  \citenamefont {Casertano}, \citenamefont {Yuan}, \citenamefont {Bowers},
  \citenamefont {Macri}, \citenamefont {Zinn},\ and\ \citenamefont
  {Scolnic}}]{Riess:2020fzl}%
  \BibitemOpen
  \bibfield  {author} {\bibinfo {author} {\bibfnamefont {A.~G.}\ \bibnamefont
  {Riess}}, \bibinfo {author} {\bibfnamefont {S.}~\bibnamefont {Casertano}},
  \bibinfo {author} {\bibfnamefont {W.}~\bibnamefont {Yuan}}, \bibinfo {author}
  {\bibfnamefont {J.~B.}\ \bibnamefont {Bowers}}, \bibinfo {author}
  {\bibfnamefont {L.}~\bibnamefont {Macri}}, \bibinfo {author} {\bibfnamefont
  {J.~C.}\ \bibnamefont {Zinn}}, \ and\ \bibinfo {author} {\bibfnamefont
  {D.}~\bibnamefont {Scolnic}},\ }\href {\doibase 10.3847/2041-8213/abdbaf}
  {\bibfield  {journal} {\bibinfo  {journal} {Astrophys. J. Lett.}\ }\textbf
  {\bibinfo {volume} {908}},\ \bibinfo {pages} {L6} (\bibinfo {year} {2021})},\
  \Eprint {http://arxiv.org/abs/2012.08534} {arXiv:2012.08534 [astro-ph.CO]}
  \BibitemShut {NoStop}%
\bibitem [{\citenamefont {Di~Valentino}\ \emph {et~al.}(2021)\citenamefont
  {Di~Valentino}, \citenamefont {Mena}, \citenamefont {Pan}, \citenamefont
  {Visinelli}, \citenamefont {Yang}, \citenamefont {Melchiorri}, \citenamefont
  {Mota}, \citenamefont {Riess},\ and\ \citenamefont
  {Silk}}]{DiValentino:2021izs}%
  \BibitemOpen
  \bibfield  {author} {\bibinfo {author} {\bibfnamefont {E.}~\bibnamefont
  {Di~Valentino}}, \bibinfo {author} {\bibfnamefont {O.}~\bibnamefont {Mena}},
  \bibinfo {author} {\bibfnamefont {S.}~\bibnamefont {Pan}}, \bibinfo {author}
  {\bibfnamefont {L.}~\bibnamefont {Visinelli}}, \bibinfo {author}
  {\bibfnamefont {W.}~\bibnamefont {Yang}}, \bibinfo {author} {\bibfnamefont
  {A.}~\bibnamefont {Melchiorri}}, \bibinfo {author} {\bibfnamefont {D.~F.}\
  \bibnamefont {Mota}}, \bibinfo {author} {\bibfnamefont {A.~G.}\ \bibnamefont
  {Riess}}, \ and\ \bibinfo {author} {\bibfnamefont {J.}~\bibnamefont {Silk}},\
  }\href {\doibase 10.1088/1361-6382/ac086d} {\  (\bibinfo {year} {2021}),\
  10.1088/1361-6382/ac086d},\ \Eprint {http://arxiv.org/abs/2103.01183}
  {arXiv:2103.01183 [astro-ph.CO]} \BibitemShut {NoStop}%
\bibitem [{\citenamefont {Abbott}\ \emph {et~al.}(2021)\citenamefont {Abbott}
  \emph {et~al.}}]{Abbott:2021bzy}%
  \BibitemOpen
  \bibfield  {author} {\bibinfo {author} {\bibfnamefont {T.~M.~C.}\
  \bibnamefont {Abbott}} \emph {et~al.} (\bibinfo {collaboration} {DES}),\
  }\href@noop {} {\  (\bibinfo {year} {2021})},\ \Eprint
  {http://arxiv.org/abs/2105.13549} {arXiv:2105.13549 [astro-ph.CO]}
  \BibitemShut {NoStop}%
\bibitem [{\citenamefont {Asgari}\ \emph {et~al.}(2021)\citenamefont {Asgari}
  \emph {et~al.}}]{Asgari:2020wuj}%
  \BibitemOpen
  \bibfield  {author} {\bibinfo {author} {\bibfnamefont {M.}~\bibnamefont
  {Asgari}} \emph {et~al.} (\bibinfo {collaboration} {KiDS}),\ }\href {\doibase
  10.1051/0004-6361/202039070} {\bibfield  {journal} {\bibinfo  {journal}
  {Astron. Astrophys.}\ }\textbf {\bibinfo {volume} {645}},\ \bibinfo {pages}
  {A104} (\bibinfo {year} {2021})},\ \Eprint {http://arxiv.org/abs/2007.15633}
  {arXiv:2007.15633 [astro-ph.CO]} \BibitemShut {NoStop}%
\bibitem [{\citenamefont {Hobson}\ \emph {et~al.}(2002)\citenamefont {Hobson},
  \citenamefont {Bridle},\ and\ \citenamefont {Lahav}}]{Hobson:2002zf}%
  \BibitemOpen
  \bibfield  {author} {\bibinfo {author} {\bibfnamefont {M.~P.}\ \bibnamefont
  {Hobson}}, \bibinfo {author} {\bibfnamefont {S.~L.}\ \bibnamefont {Bridle}},
  \ and\ \bibinfo {author} {\bibfnamefont {O.}~\bibnamefont {Lahav}},\ }\href
  {\doibase 10.1046/j.1365-8711.2002.05614.x} {\bibfield  {journal} {\bibinfo
  {journal} {Mon. Not. Roy. Astron. Soc.}\ }\textbf {\bibinfo {volume} {335}},\
  \bibinfo {pages} {377} (\bibinfo {year} {2002})},\ \Eprint
  {http://arxiv.org/abs/astro-ph/0203259} {arXiv:astro-ph/0203259} \BibitemShut
  {NoStop}%
\bibitem [{\citenamefont {Marshall}\ \emph {et~al.}(2006)\citenamefont
  {Marshall}, \citenamefont {Rajguru},\ and\ \citenamefont
  {Slosar}}]{Marshall:2004zd}%
  \BibitemOpen
  \bibfield  {author} {\bibinfo {author} {\bibfnamefont {P.}~\bibnamefont
  {Marshall}}, \bibinfo {author} {\bibfnamefont {N.}~\bibnamefont {Rajguru}}, \
  and\ \bibinfo {author} {\bibfnamefont {A.}~\bibnamefont {Slosar}},\ }\href
  {\doibase 10.1103/PhysRevD.73.067302} {\bibfield  {journal} {\bibinfo
  {journal} {Phys. Rev. D}\ }\textbf {\bibinfo {volume} {73}},\ \bibinfo
  {pages} {067302} (\bibinfo {year} {2006})},\ \Eprint
  {http://arxiv.org/abs/astro-ph/0412535} {arXiv:astro-ph/0412535} \BibitemShut
  {NoStop}%
\bibitem [{\citenamefont {Amendola}\ \emph {et~al.}(2013)\citenamefont
  {Amendola}, \citenamefont {Marra},\ and\ \citenamefont
  {Quartin}}]{Amendola:2012wc}%
  \BibitemOpen
  \bibfield  {author} {\bibinfo {author} {\bibfnamefont {L.}~\bibnamefont
  {Amendola}}, \bibinfo {author} {\bibfnamefont {V.}~\bibnamefont {Marra}}, \
  and\ \bibinfo {author} {\bibfnamefont {M.}~\bibnamefont {Quartin}},\ }\href
  {\doibase 10.1093/mnras/stt008} {\bibfield  {journal} {\bibinfo  {journal}
  {Mon. Not. Roy. Astron. Soc.}\ }\textbf {\bibinfo {volume} {430}},\ \bibinfo
  {pages} {1867} (\bibinfo {year} {2013})},\ \Eprint
  {http://arxiv.org/abs/1209.1897} {arXiv:1209.1897 [astro-ph.CO]} \BibitemShut
  {NoStop}%
\bibitem [{\citenamefont {Martin}\ \emph {et~al.}(2014)\citenamefont {Martin},
  \citenamefont {Ringeval}, \citenamefont {Trotta},\ and\ \citenamefont
  {Vennin}}]{Martin:2014lra}%
  \BibitemOpen
  \bibfield  {author} {\bibinfo {author} {\bibfnamefont {J.}~\bibnamefont
  {Martin}}, \bibinfo {author} {\bibfnamefont {C.}~\bibnamefont {Ringeval}},
  \bibinfo {author} {\bibfnamefont {R.}~\bibnamefont {Trotta}}, \ and\ \bibinfo
  {author} {\bibfnamefont {V.}~\bibnamefont {Vennin}},\ }\href {\doibase
  10.1103/PhysRevD.90.063501} {\bibfield  {journal} {\bibinfo  {journal} {Phys.
  Rev. D}\ }\textbf {\bibinfo {volume} {90}},\ \bibinfo {pages} {063501}
  (\bibinfo {year} {2014})},\ \Eprint {http://arxiv.org/abs/1405.7272}
  {arXiv:1405.7272 [astro-ph.CO]} \BibitemShut {NoStop}%
\bibitem [{\citenamefont {Raveri}(2016)}]{Raveri:2015maa}%
  \BibitemOpen
  \bibfield  {author} {\bibinfo {author} {\bibfnamefont {M.}~\bibnamefont
  {Raveri}},\ }\href {\doibase 10.1103/PhysRevD.93.043522} {\bibfield
  {journal} {\bibinfo  {journal} {Phys. Rev. D}\ }\textbf {\bibinfo {volume}
  {93}},\ \bibinfo {pages} {043522} (\bibinfo {year} {2016})},\ \Eprint
  {http://arxiv.org/abs/1510.00688} {arXiv:1510.00688 [astro-ph.CO]}
  \BibitemShut {NoStop}%
\bibitem [{\citenamefont {Seehars}\ \emph {et~al.}(2016)\citenamefont
  {Seehars}, \citenamefont {Grandis}, \citenamefont {Amara},\ and\
  \citenamefont {Refregier}}]{Seehars:2015qza}%
  \BibitemOpen
  \bibfield  {author} {\bibinfo {author} {\bibfnamefont {S.}~\bibnamefont
  {Seehars}}, \bibinfo {author} {\bibfnamefont {S.}~\bibnamefont {Grandis}},
  \bibinfo {author} {\bibfnamefont {A.}~\bibnamefont {Amara}}, \ and\ \bibinfo
  {author} {\bibfnamefont {A.}~\bibnamefont {Refregier}},\ }\href {\doibase
  10.1103/PhysRevD.93.103507} {\bibfield  {journal} {\bibinfo  {journal} {Phys.
  Rev. D}\ }\textbf {\bibinfo {volume} {93}},\ \bibinfo {pages} {103507}
  (\bibinfo {year} {2016})},\ \Eprint {http://arxiv.org/abs/1510.08483}
  {arXiv:1510.08483 [astro-ph.CO]} \BibitemShut {NoStop}%
\bibitem [{\citenamefont {Joudaki}\ \emph {et~al.}(2017)\citenamefont {Joudaki}
  \emph {et~al.}}]{Joudaki:2016mvz}%
  \BibitemOpen
  \bibfield  {author} {\bibinfo {author} {\bibfnamefont {S.}~\bibnamefont
  {Joudaki}} \emph {et~al.},\ }\href {\doibase 10.1093/mnras/stw2665}
  {\bibfield  {journal} {\bibinfo  {journal} {Mon. Not. Roy. Astron. Soc.}\
  }\textbf {\bibinfo {volume} {465}},\ \bibinfo {pages} {2033} (\bibinfo {year}
  {2017})},\ \Eprint {http://arxiv.org/abs/1601.05786} {arXiv:1601.05786
  [astro-ph.CO]} \BibitemShut {NoStop}%
\bibitem [{\citenamefont {Charnock}\ \emph {et~al.}(2017)\citenamefont
  {Charnock}, \citenamefont {Battye},\ and\ \citenamefont
  {Moss}}]{Charnock:2017vcd}%
  \BibitemOpen
  \bibfield  {author} {\bibinfo {author} {\bibfnamefont {T.}~\bibnamefont
  {Charnock}}, \bibinfo {author} {\bibfnamefont {R.~A.}\ \bibnamefont
  {Battye}}, \ and\ \bibinfo {author} {\bibfnamefont {A.}~\bibnamefont
  {Moss}},\ }\href {\doibase 10.1103/PhysRevD.95.123535} {\bibfield  {journal}
  {\bibinfo  {journal} {Phys. Rev. D}\ }\textbf {\bibinfo {volume} {95}},\
  \bibinfo {pages} {123535} (\bibinfo {year} {2017})},\ \Eprint
  {http://arxiv.org/abs/1703.05959} {arXiv:1703.05959 [astro-ph.CO]}
  \BibitemShut {NoStop}%
\bibitem [{\citenamefont {Lin}\ and\ \citenamefont
  {Ishak}(2017)}]{Lin:2017ikq}%
  \BibitemOpen
  \bibfield  {author} {\bibinfo {author} {\bibfnamefont {W.}~\bibnamefont
  {Lin}}\ and\ \bibinfo {author} {\bibfnamefont {M.}~\bibnamefont {Ishak}},\
  }\href {\doibase 10.1103/PhysRevD.96.023532} {\bibfield  {journal} {\bibinfo
  {journal} {Phys. Rev. D}\ }\textbf {\bibinfo {volume} {96}},\ \bibinfo
  {pages} {023532} (\bibinfo {year} {2017})},\ \Eprint
  {http://arxiv.org/abs/1705.05303} {arXiv:1705.05303 [astro-ph.CO]}
  \BibitemShut {NoStop}%
\bibitem [{\citenamefont {Luis~Bernal}\ and\ \citenamefont
  {Peacock}(2018)}]{Bernal:2018cxc}%
  \BibitemOpen
  \bibfield  {author} {\bibinfo {author} {\bibfnamefont {J.}~\bibnamefont
  {Luis~Bernal}}\ and\ \bibinfo {author} {\bibfnamefont {J.~A.}\ \bibnamefont
  {Peacock}},\ }\href {\doibase 10.1088/1475-7516/2018/07/002} {\bibfield
  {journal} {\bibinfo  {journal} {JCAP}\ }\textbf {\bibinfo {volume} {07}},\
  \bibinfo {pages} {002} (\bibinfo {year} {2018})},\ \Eprint
  {http://arxiv.org/abs/1803.04470} {arXiv:1803.04470 [astro-ph.CO]}
  \BibitemShut {NoStop}%
\bibitem [{\citenamefont {Raveri}\ and\ \citenamefont
  {Hu}(2019)}]{Raveri:2018wln}%
  \BibitemOpen
  \bibfield  {author} {\bibinfo {author} {\bibfnamefont {M.}~\bibnamefont
  {Raveri}}\ and\ \bibinfo {author} {\bibfnamefont {W.}~\bibnamefont {Hu}},\
  }\href {\doibase 10.1103/PhysRevD.99.043506} {\bibfield  {journal} {\bibinfo
  {journal} {Phys. Rev. D}\ }\textbf {\bibinfo {volume} {99}},\ \bibinfo
  {pages} {043506} (\bibinfo {year} {2019})},\ \Eprint
  {http://arxiv.org/abs/1806.04649} {arXiv:1806.04649 [astro-ph.CO]}
  \BibitemShut {NoStop}%
\bibitem [{\citenamefont {Adhikari}\ and\ \citenamefont
  {Huterer}(2019)}]{Adhikari:2018wnk}%
  \BibitemOpen
  \bibfield  {author} {\bibinfo {author} {\bibfnamefont {S.}~\bibnamefont
  {Adhikari}}\ and\ \bibinfo {author} {\bibfnamefont {D.}~\bibnamefont
  {Huterer}},\ }\href {\doibase 10.1088/1475-7516/2019/01/036} {\bibfield
  {journal} {\bibinfo  {journal} {JCAP}\ }\textbf {\bibinfo {volume} {01}},\
  \bibinfo {pages} {036} (\bibinfo {year} {2019})},\ \Eprint
  {http://arxiv.org/abs/1806.04292} {arXiv:1806.04292 [astro-ph.CO]}
  \BibitemShut {NoStop}%
\bibitem [{\citenamefont {Raveri}\ \emph {et~al.}(2020)\citenamefont {Raveri},
  \citenamefont {Zacharegkas},\ and\ \citenamefont {Hu}}]{Raveri:2019gdp}%
  \BibitemOpen
  \bibfield  {author} {\bibinfo {author} {\bibfnamefont {M.}~\bibnamefont
  {Raveri}}, \bibinfo {author} {\bibfnamefont {G.}~\bibnamefont {Zacharegkas}},
  \ and\ \bibinfo {author} {\bibfnamefont {W.}~\bibnamefont {Hu}},\ }\href
  {\doibase 10.1103/PhysRevD.101.103527} {\bibfield  {journal} {\bibinfo
  {journal} {Phys. Rev. D}\ }\textbf {\bibinfo {volume} {101}},\ \bibinfo
  {pages} {103527} (\bibinfo {year} {2020})},\ \Eprint
  {http://arxiv.org/abs/1912.04880} {arXiv:1912.04880 [astro-ph.CO]}
  \BibitemShut {NoStop}%
\bibitem [{\citenamefont {Handley}\ and\ \citenamefont
  {Lemos}(2019)}]{Handley:2019wlz}%
  \BibitemOpen
  \bibfield  {author} {\bibinfo {author} {\bibfnamefont {W.}~\bibnamefont
  {Handley}}\ and\ \bibinfo {author} {\bibfnamefont {P.}~\bibnamefont
  {Lemos}},\ }\href {\doibase 10.1103/PhysRevD.100.043504} {\bibfield
  {journal} {\bibinfo  {journal} {Phys. Rev. D}\ }\textbf {\bibinfo {volume}
  {100}},\ \bibinfo {pages} {043504} (\bibinfo {year} {2019})},\ \Eprint
  {http://arxiv.org/abs/1902.04029} {arXiv:1902.04029 [astro-ph.CO]}
  \BibitemShut {NoStop}%
\bibitem [{\citenamefont {Lemos}\ \emph
  {et~al.}(2020{\natexlab{a}})\citenamefont {Lemos}, \citenamefont
  {K\"ohlinger}, \citenamefont {Handley}, \citenamefont {Joachimi},
  \citenamefont {Whiteway},\ and\ \citenamefont {Lahav}}]{Lemos:2019txn}%
  \BibitemOpen
  \bibfield  {author} {\bibinfo {author} {\bibfnamefont {P.}~\bibnamefont
  {Lemos}}, \bibinfo {author} {\bibfnamefont {F.}~\bibnamefont {K\"ohlinger}},
  \bibinfo {author} {\bibfnamefont {W.}~\bibnamefont {Handley}}, \bibinfo
  {author} {\bibfnamefont {B.}~\bibnamefont {Joachimi}}, \bibinfo {author}
  {\bibfnamefont {L.}~\bibnamefont {Whiteway}}, \ and\ \bibinfo {author}
  {\bibfnamefont {O.}~\bibnamefont {Lahav}},\ }\href {\doibase
  10.1093/mnras/staa1836} {\bibfield  {journal} {\bibinfo  {journal} {Mon. Not.
  Roy. Astron. Soc.}\ }\textbf {\bibinfo {volume} {496}},\ \bibinfo {pages}
  {4647} (\bibinfo {year} {2020}{\natexlab{a}})},\ \Eprint
  {http://arxiv.org/abs/1910.07820} {arXiv:1910.07820 [astro-ph.CO]}
  \BibitemShut {NoStop}%
\bibitem [{\citenamefont {Lin}\ and\ \citenamefont
  {Ishak}(2019)}]{Lin:2019zdn}%
  \BibitemOpen
  \bibfield  {author} {\bibinfo {author} {\bibfnamefont {W.}~\bibnamefont
  {Lin}}\ and\ \bibinfo {author} {\bibfnamefont {M.}~\bibnamefont {Ishak}},\
  }\href@noop {} {\  (\bibinfo {year} {2019})},\ \Eprint
  {http://arxiv.org/abs/1909.10991} {arXiv:1909.10991 [astro-ph.CO]}
  \BibitemShut {NoStop}%
\bibitem [{\citenamefont {Park}\ and\ \citenamefont
  {Rozo}(2020)}]{Park:2019tyw}%
  \BibitemOpen
  \bibfield  {author} {\bibinfo {author} {\bibfnamefont {Y.}~\bibnamefont
  {Park}}\ and\ \bibinfo {author} {\bibfnamefont {E.}~\bibnamefont {Rozo}},\
  }\href {\doibase 10.1093/mnras/staa2647} {\bibfield  {journal} {\bibinfo
  {journal} {Mon. Not. Roy. Astron. Soc.}\ }\textbf {\bibinfo {volume} {499}},\
  \bibinfo {pages} {4638} (\bibinfo {year} {2020})},\ \Eprint
  {http://arxiv.org/abs/1907.05798} {arXiv:1907.05798 [astro-ph.CO]}
  \BibitemShut {NoStop}%
\bibitem [{\citenamefont {Miranda}\ \emph {et~al.}(2020)\citenamefont
  {Miranda}, \citenamefont {Rogozenski},\ and\ \citenamefont
  {Krause}}]{Miranda:2020lpk}%
  \BibitemOpen
  \bibfield  {author} {\bibinfo {author} {\bibfnamefont {V.}~\bibnamefont
  {Miranda}}, \bibinfo {author} {\bibfnamefont {P.}~\bibnamefont {Rogozenski}},
  \ and\ \bibinfo {author} {\bibfnamefont {E.}~\bibnamefont {Krause}},\
  }\href@noop {} {\  (\bibinfo {year} {2020})},\ \Eprint
  {http://arxiv.org/abs/2009.14241} {arXiv:2009.14241 [astro-ph.CO]}
  \BibitemShut {NoStop}%
\bibitem [{\citenamefont {Doux}\ \emph {et~al.}(2021)\citenamefont {Doux} \emph
  {et~al.}}]{Doux:2020kdz}%
  \BibitemOpen
  \bibfield  {author} {\bibinfo {author} {\bibfnamefont {C.}~\bibnamefont
  {Doux}} \emph {et~al.} (\bibinfo {collaboration} {DES}),\ }\href {\doibase
  10.1093/mnras/stab526} {\bibfield  {journal} {\bibinfo  {journal} {Mon. Not.
  Roy. Astron. Soc.}\ }\textbf {\bibinfo {volume} {503}},\ \bibinfo {pages}
  {2688} (\bibinfo {year} {2021})},\ \Eprint {http://arxiv.org/abs/2011.03410}
  {arXiv:2011.03410 [astro-ph.CO]} \BibitemShut {NoStop}%
\bibitem [{\citenamefont {Lemos}\ \emph
  {et~al.}(2020{\natexlab{b}})\citenamefont {Lemos} \emph
  {et~al.}}]{Lemos:2020jry}%
  \BibitemOpen
  \bibfield  {author} {\bibinfo {author} {\bibfnamefont {P.}~\bibnamefont
  {Lemos}} \emph {et~al.} (\bibinfo {collaboration} {DES}),\ }\href@noop {} {\
  (\bibinfo {year} {2020}{\natexlab{b}})},\ \Eprint
  {http://arxiv.org/abs/2012.09554} {arXiv:2012.09554 [astro-ph.CO]}
  \BibitemShut {NoStop}%
\bibitem [{\citenamefont {Abbott}\ \emph {et~al.}(2018)\citenamefont {Abbott}
  \emph {et~al.}}]{Abbott:2017wau}%
  \BibitemOpen
  \bibfield  {author} {\bibinfo {author} {\bibfnamefont {T.~M.~C.}\
  \bibnamefont {Abbott}} \emph {et~al.} (\bibinfo {collaboration} {DES}),\
  }\href {\doibase 10.1103/PhysRevD.98.043526} {\bibfield  {journal} {\bibinfo
  {journal} {Phys. Rev. D}\ }\textbf {\bibinfo {volume} {98}},\ \bibinfo
  {pages} {043526} (\bibinfo {year} {2018})},\ \Eprint
  {http://arxiv.org/abs/1708.01530} {arXiv:1708.01530 [astro-ph.CO]}
  \BibitemShut {NoStop}%
\bibitem [{\citenamefont {Wand}\ and\ \citenamefont
  {Jones}(1994)}]{wand1994kernel}%
  \BibitemOpen
  \bibfield  {author} {\bibinfo {author} {\bibfnamefont {M.}~\bibnamefont
  {Wand}}\ and\ \bibinfo {author} {\bibfnamefont {M.}~\bibnamefont {Jones}},\
  }\href@noop {} {\emph {\bibinfo {title} {Kernel Smoothing}}},\ Chapman \&
  Hall/CRC Monographs on Statistics \& Applied Probability\ (\bibinfo
  {publisher} {Taylor \& Francis},\ \bibinfo {year} {1994})\BibitemShut
  {NoStop}%
\bibitem [{\citenamefont {Chac{\'o}n}\ and\ \citenamefont
  {Duong}(2018)}]{chacon2018multivariate}%
  \BibitemOpen
  \bibfield  {author} {\bibinfo {author} {\bibfnamefont {J.}~\bibnamefont
  {Chac{\'o}n}}\ and\ \bibinfo {author} {\bibfnamefont {T.}~\bibnamefont
  {Duong}},\ }\href@noop {} {\emph {\bibinfo {title} {Multivariate Kernel
  Smoothing and Its Applications}}},\ Chapman \& Hall/CRC Monographs on
  Statistics and Applied Probability\ (\bibinfo  {publisher} {CRC Press},\
  \bibinfo {year} {2018})\BibitemShut {NoStop}%
\bibitem [{\citenamefont {Terrell}\ and\ \citenamefont
  {Scott}(1992)}]{terrell1992}%
  \BibitemOpen
  \bibfield  {author} {\bibinfo {author} {\bibfnamefont {G.~R.}\ \bibnamefont
  {Terrell}}\ and\ \bibinfo {author} {\bibfnamefont {D.~W.}\ \bibnamefont
  {Scott}},\ }\href {\doibase 10.1214/aos/1176348768} {\bibfield  {journal}
  {\bibinfo  {journal} {Ann. Statist.}\ }\textbf {\bibinfo {volume} {20}},\
  \bibinfo {pages} {1236} (\bibinfo {year} {1992})}\BibitemShut {NoStop}%
\bibitem [{\citenamefont {Rosenblatt}(1956)}]{Rosenblatt:1956:RSN}%
  \BibitemOpen
  \bibfield  {author} {\bibinfo {author} {\bibfnamefont {M.}~\bibnamefont
  {Rosenblatt}},\ }\href {\doibase https://doi.org/10.1214/aoms/1177728190}
  {\bibfield  {journal} {\bibinfo  {journal} {j-ANN-MATH-STAT}\ }\textbf
  {\bibinfo {volume} {27}},\ \bibinfo {pages} {832} (\bibinfo {year}
  {1956})}\BibitemShut {NoStop}%
\bibitem [{\citenamefont {Parzen}(1962)}]{parzen1962}%
  \BibitemOpen
  \bibfield  {author} {\bibinfo {author} {\bibfnamefont {E.}~\bibnamefont
  {Parzen}},\ }\href {\doibase 10.1214/aoms/1177704472} {\bibfield  {journal}
  {\bibinfo  {journal} {Ann. Math. Statist.}\ }\textbf {\bibinfo {volume}
  {33}},\ \bibinfo {pages} {1065} (\bibinfo {year} {1962})}\BibitemShut
  {NoStop}%
\bibitem [{\citenamefont {Bagnato}\ and\ \citenamefont
  {Punzo}(2020)}]{2019arXiv190600587B}%
  \BibitemOpen
  \bibfield  {author} {\bibinfo {author} {\bibfnamefont {L.}~\bibnamefont
  {Bagnato}}\ and\ \bibinfo {author} {\bibfnamefont {A.}~\bibnamefont
  {Punzo}},\ }\href {\doibase 10.1007/s00180-020-01041-8} {\bibfield  {journal}
  {\bibinfo  {journal} {Computational Statistics}\ } (\bibinfo {year} {2020}),\
  10.1007/s00180-020-01041-8}\BibitemShut {NoStop}%
\bibitem [{\citenamefont {Papamakarios}\ \emph {et~al.}(2017)\citenamefont
  {Papamakarios}, \citenamefont {Pavlakou},\ and\ \citenamefont
  {Murray}}]{2017arXiv170507057P}%
  \BibitemOpen
  \bibfield  {author} {\bibinfo {author} {\bibfnamefont {G.}~\bibnamefont
  {Papamakarios}}, \bibinfo {author} {\bibfnamefont {T.}~\bibnamefont
  {Pavlakou}}, \ and\ \bibinfo {author} {\bibfnamefont {I.}~\bibnamefont
  {Murray}},\ }\href@noop {} {\  (\bibinfo {year} {2017})},\ \Eprint
  {http://arxiv.org/abs/1705.07057} {arXiv:1705.07057 [stat.ML]} \BibitemShut
  {NoStop}%
\bibitem [{\citenamefont {Kingma}\ and\ \citenamefont
  {Dhariwal}(2018)}]{2018arXiv180703039K}%
  \BibitemOpen
  \bibfield  {author} {\bibinfo {author} {\bibfnamefont {D.~P.}\ \bibnamefont
  {Kingma}}\ and\ \bibinfo {author} {\bibfnamefont {P.}~\bibnamefont
  {Dhariwal}},\ }\href@noop {} {\  (\bibinfo {year} {2018})},\ \Eprint
  {http://arxiv.org/abs/1807.03039} {arXiv:1807.03039 [stat.ML]} \BibitemShut
  {NoStop}%
\bibitem [{\citenamefont {Grathwohl}\ \emph {et~al.}(2018)\citenamefont
  {Grathwohl}, \citenamefont {Chen}, \citenamefont {Bettencourt}, \citenamefont
  {Sutskever},\ and\ \citenamefont {Duvenaud}}]{2018arXiv181001367G}%
  \BibitemOpen
  \bibfield  {author} {\bibinfo {author} {\bibfnamefont {W.}~\bibnamefont
  {Grathwohl}}, \bibinfo {author} {\bibfnamefont {R.~T.~Q.}\ \bibnamefont
  {Chen}}, \bibinfo {author} {\bibfnamefont {J.}~\bibnamefont {Bettencourt}},
  \bibinfo {author} {\bibfnamefont {I.}~\bibnamefont {Sutskever}}, \ and\
  \bibinfo {author} {\bibfnamefont {D.}~\bibnamefont {Duvenaud}},\ }\href@noop
  {} {\  (\bibinfo {year} {2018})},\ \Eprint {http://arxiv.org/abs/1810.01367}
  {arXiv:1810.01367 [cs.LG]} \BibitemShut {NoStop}%
\bibitem [{\citenamefont {Kobyzev}\ \emph {et~al.}(2019)\citenamefont
  {Kobyzev}, \citenamefont {Prince},\ and\ \citenamefont
  {Brubaker}}]{Kobyzev:2019fa}%
  \BibitemOpen
  \bibfield  {author} {\bibinfo {author} {\bibfnamefont {I.}~\bibnamefont
  {Kobyzev}}, \bibinfo {author} {\bibfnamefont {S.~J.~D.}\ \bibnamefont
  {Prince}}, \ and\ \bibinfo {author} {\bibfnamefont {M.~A.}\ \bibnamefont
  {Brubaker}},\ }\href {\doibase 10.1109/TPAMI.2020.2992934} {\bibfield
  {journal} {\bibinfo  {journal} {IEEE Trans. Pattern Anal. Mach. Intell.}\ ,\
  \bibinfo {pages} {1}} (\bibinfo {year} {2019})},\ \Eprint
  {http://arxiv.org/abs/1908.09257} {1908.09257} \BibitemShut {NoStop}%
\bibitem [{\citenamefont {Meng}\ \emph {et~al.}(2020)\citenamefont {Meng},
  \citenamefont {Song}, \citenamefont {Song},\ and\ \citenamefont
  {Ermon}}]{2020arXiv200301941M}%
  \BibitemOpen
  \bibfield  {author} {\bibinfo {author} {\bibfnamefont {C.}~\bibnamefont
  {Meng}}, \bibinfo {author} {\bibfnamefont {Y.}~\bibnamefont {Song}}, \bibinfo
  {author} {\bibfnamefont {J.}~\bibnamefont {Song}}, \ and\ \bibinfo {author}
  {\bibfnamefont {S.}~\bibnamefont {Ermon}},\ }\href@noop {} {\  (\bibinfo
  {year} {2020})},\ \Eprint {http://arxiv.org/abs/2003.01941} {arXiv:2003.01941
  [cs.LG]} \BibitemShut {NoStop}%
\bibitem [{\citenamefont {Reiman}\ \emph {et~al.}(2020)\citenamefont {Reiman},
  \citenamefont {Tamanas}, \citenamefont {Prochaska},\ and\ \citenamefont
  {\v{D}urov\v{c}\'\i{}kov\'a}}]{2020arXiv200600615R}%
  \BibitemOpen
  \bibfield  {author} {\bibinfo {author} {\bibfnamefont {D.~M.}\ \bibnamefont
  {Reiman}}, \bibinfo {author} {\bibfnamefont {J.}~\bibnamefont {Tamanas}},
  \bibinfo {author} {\bibfnamefont {J.~X.}\ \bibnamefont {Prochaska}}, \ and\
  \bibinfo {author} {\bibfnamefont {D.}~\bibnamefont
  {\v{D}urov\v{c}\'\i{}kov\'a}},\ }\href@noop {} {\  (\bibinfo {year}
  {2020})},\ \Eprint {http://arxiv.org/abs/2006.00615} {arXiv:2006.00615
  [astro-ph.CO]} \BibitemShut {NoStop}%
\bibitem [{\citenamefont {Diaz~Rivero}\ and\ \citenamefont
  {Dvorkin}(2020)}]{2020PhRvD.102j3507D}%
  \BibitemOpen
  \bibfield  {author} {\bibinfo {author} {\bibfnamefont {A.}~\bibnamefont
  {Diaz~Rivero}}\ and\ \bibinfo {author} {\bibfnamefont {C.}~\bibnamefont
  {Dvorkin}},\ }\href {\doibase 10.1103/PhysRevD.102.103507} {\bibfield
  {journal} {\bibinfo  {journal} {Phys. Rev. D}\ }\textbf {\bibinfo {volume}
  {102}},\ \bibinfo {pages} {103507} (\bibinfo {year} {2020})},\ \Eprint
  {http://arxiv.org/abs/2007.05535} {arXiv:2007.05535 [astro-ph.CO]}
  \BibitemShut {NoStop}%
\bibitem [{\citenamefont {Moss}(2020)}]{2020MNRAS.496..328M}%
  \BibitemOpen
  \bibfield  {author} {\bibinfo {author} {\bibfnamefont {A.}~\bibnamefont
  {Moss}},\ }\href {\doibase 10.1093/mnras/staa1469} {\bibfield  {journal}
  {\bibinfo  {journal} {Mon. Not. Roy. Astron. Soc.}\ }\textbf {\bibinfo
  {volume} {496}},\ \bibinfo {pages} {328} (\bibinfo {year} {2020})},\ \Eprint
  {http://arxiv.org/abs/1903.10860} {arXiv:1903.10860 [astro-ph.CO]}
  \BibitemShut {NoStop}%
\bibitem [{\citenamefont {Jeffrey}\ \emph {et~al.}(2021)\citenamefont
  {Jeffrey}, \citenamefont {Alsing},\ and\ \citenamefont
  {Lanusse}}]{2020MNRAS.tmp.3445J}%
  \BibitemOpen
  \bibfield  {author} {\bibinfo {author} {\bibfnamefont {N.}~\bibnamefont
  {Jeffrey}}, \bibinfo {author} {\bibfnamefont {J.}~\bibnamefont {Alsing}}, \
  and\ \bibinfo {author} {\bibfnamefont {F.}~\bibnamefont {Lanusse}},\ }\href
  {\doibase 10.1093/mnras/staa3594} {\bibfield  {journal} {\bibinfo  {journal}
  {Mon. Not. Roy. Astron. Soc.}\ }\textbf {\bibinfo {volume} {501}},\ \bibinfo
  {pages} {954} (\bibinfo {year} {2021})},\ \Eprint
  {http://arxiv.org/abs/2009.08459} {arXiv:2009.08459 [astro-ph.CO]}
  \BibitemShut {NoStop}%
\bibitem [{\citenamefont {Bengio}\ and\ \citenamefont
  {Lecun}(2007)}]{bengio:2007}%
  \BibitemOpen
  \bibfield  {author} {\bibinfo {author} {\bibfnamefont {Y.}~\bibnamefont
  {Bengio}}\ and\ \bibinfo {author} {\bibfnamefont {Y.}~\bibnamefont {Lecun}},\
  }\enquote {\bibinfo {title} {Scaling learning algorithms towards ai},}\ in\
  \href@noop {} {\emph {\bibinfo {booktitle} {Large-scale kernel machines}}},\
  \bibinfo {editor} {edited by\ \bibinfo {editor} {\bibfnamefont
  {L.}~\bibnamefont {Bottou}}, \bibinfo {editor} {\bibfnamefont
  {O.}~\bibnamefont {Chapelle}}, \bibinfo {editor} {\bibfnamefont
  {D.}~\bibnamefont {DeCoste}}, \ and\ \bibinfo {editor} {\bibfnamefont
  {J.}~\bibnamefont {Weston}}}\ (\bibinfo  {publisher} {MIT Press},\ \bibinfo
  {year} {2007})\BibitemShut {NoStop}%
\bibitem [{\citenamefont {Germain}\ \emph {et~al.}(2015)\citenamefont
  {Germain}, \citenamefont {Gregor}, \citenamefont {Murray},\ and\
  \citenamefont {Larochelle}}]{2015arXiv150203509G}%
  \BibitemOpen
  \bibfield  {author} {\bibinfo {author} {\bibfnamefont {M.}~\bibnamefont
  {Germain}}, \bibinfo {author} {\bibfnamefont {K.}~\bibnamefont {Gregor}},
  \bibinfo {author} {\bibfnamefont {I.}~\bibnamefont {Murray}}, \ and\ \bibinfo
  {author} {\bibfnamefont {H.}~\bibnamefont {Larochelle}},\ }\href@noop {} {\
  (\bibinfo {year} {2015})},\ \Eprint {http://arxiv.org/abs/1502.03509}
  {arXiv:1502.03509 [cs.LG]} \BibitemShut {NoStop}%
\bibitem [{\citenamefont {Dillon}\ \emph {et~al.}(2017)\citenamefont {Dillon},
  \citenamefont {Langmore}, \citenamefont {Tran}, \citenamefont {Brevdo},
  \citenamefont {Vasudevan}, \citenamefont {Moore}, \citenamefont {Patton},
  \citenamefont {Alemi}, \citenamefont {Hoffman},\ and\ \citenamefont
  {Saurous}}]{2017arXiv171110604D}%
  \BibitemOpen
  \bibfield  {author} {\bibinfo {author} {\bibfnamefont {J.~V.}\ \bibnamefont
  {Dillon}}, \bibinfo {author} {\bibfnamefont {I.}~\bibnamefont {Langmore}},
  \bibinfo {author} {\bibfnamefont {D.}~\bibnamefont {Tran}}, \bibinfo {author}
  {\bibfnamefont {E.}~\bibnamefont {Brevdo}}, \bibinfo {author} {\bibfnamefont
  {S.}~\bibnamefont {Vasudevan}}, \bibinfo {author} {\bibfnamefont
  {D.}~\bibnamefont {Moore}}, \bibinfo {author} {\bibfnamefont
  {B.}~\bibnamefont {Patton}}, \bibinfo {author} {\bibfnamefont
  {A.}~\bibnamefont {Alemi}}, \bibinfo {author} {\bibfnamefont
  {M.}~\bibnamefont {Hoffman}}, \ and\ \bibinfo {author} {\bibfnamefont
  {R.~A.}\ \bibnamefont {Saurous}},\ }\href@noop {} {\  (\bibinfo {year}
  {2017})},\ \Eprint {http://arxiv.org/abs/1711.10604} {arXiv:1711.10604
  [cs.LG]} \BibitemShut {NoStop}%
\bibitem [{\citenamefont {Zuntz}\ \emph {et~al.}(2015)\citenamefont {Zuntz},
  \citenamefont {Paterno}, \citenamefont {Jennings}, \citenamefont {Rudd},
  \citenamefont {Manzotti}, \citenamefont {Dodelson}, \citenamefont {Bridle},
  \citenamefont {Sehrish},\ and\ \citenamefont {Kowalkowski}}]{Zuntz:2014csq}%
  \BibitemOpen
  \bibfield  {author} {\bibinfo {author} {\bibfnamefont {J.}~\bibnamefont
  {Zuntz}}, \bibinfo {author} {\bibfnamefont {M.}~\bibnamefont {Paterno}},
  \bibinfo {author} {\bibfnamefont {E.}~\bibnamefont {Jennings}}, \bibinfo
  {author} {\bibfnamefont {D.}~\bibnamefont {Rudd}}, \bibinfo {author}
  {\bibfnamefont {A.}~\bibnamefont {Manzotti}}, \bibinfo {author}
  {\bibfnamefont {S.}~\bibnamefont {Dodelson}}, \bibinfo {author}
  {\bibfnamefont {S.}~\bibnamefont {Bridle}}, \bibinfo {author} {\bibfnamefont
  {S.}~\bibnamefont {Sehrish}}, \ and\ \bibinfo {author} {\bibfnamefont
  {J.}~\bibnamefont {Kowalkowski}},\ }\href {\doibase
  10.1016/j.ascom.2015.05.005} {\bibfield  {journal} {\bibinfo  {journal}
  {Astron. Comput.}\ }\textbf {\bibinfo {volume} {12}},\ \bibinfo {pages} {45}
  (\bibinfo {year} {2015})},\ \Eprint {http://arxiv.org/abs/1409.3409}
  {arXiv:1409.3409 [astro-ph.CO]} \BibitemShut {NoStop}%
\bibitem [{\citenamefont {Feroz}\ and\ \citenamefont
  {Hobson}(2008)}]{Feroz:2007kg}%
  \BibitemOpen
  \bibfield  {author} {\bibinfo {author} {\bibfnamefont {F.}~\bibnamefont
  {Feroz}}\ and\ \bibinfo {author} {\bibfnamefont {M.~P.}\ \bibnamefont
  {Hobson}},\ }\href {\doibase 10.1111/j.1365-2966.2007.12353.x} {\bibfield
  {journal} {\bibinfo  {journal} {Mon. Not. Roy. Astron. Soc.}\ }\textbf
  {\bibinfo {volume} {384}},\ \bibinfo {pages} {449} (\bibinfo {year}
  {2008})},\ \Eprint {http://arxiv.org/abs/0704.3704} {arXiv:0704.3704
  [astro-ph]} \BibitemShut {NoStop}%
\bibitem [{\citenamefont {Long}\ \emph {et~al.}(2018)\citenamefont {Long},
  \citenamefont {Raveri}, \citenamefont {Hu},\ and\ \citenamefont
  {Dodelson}}]{Long:2017dru}%
  \BibitemOpen
  \bibfield  {author} {\bibinfo {author} {\bibfnamefont {A.~J.}\ \bibnamefont
  {Long}}, \bibinfo {author} {\bibfnamefont {M.}~\bibnamefont {Raveri}},
  \bibinfo {author} {\bibfnamefont {W.}~\bibnamefont {Hu}}, \ and\ \bibinfo
  {author} {\bibfnamefont {S.}~\bibnamefont {Dodelson}},\ }\href {\doibase
  10.1103/PhysRevD.97.043510} {\bibfield  {journal} {\bibinfo  {journal} {Phys.
  Rev. D}\ }\textbf {\bibinfo {volume} {97}},\ \bibinfo {pages} {043510}
  (\bibinfo {year} {2018})},\ \Eprint {http://arxiv.org/abs/1711.08434}
  {arXiv:1711.08434 [astro-ph.CO]} \BibitemShut {NoStop}%
\end{thebibliography}%

\end{document}